\documentclass[11pt]{article}
\usepackage[]{geometry}
\usepackage{graphicx}
\usepackage{caption}
\usepackage{subcaption}
\usepackage{hyperref}
\usepackage{amsmath}
\usepackage{amsthm}
\usepackage{amssymb}
\usepackage{authblk}
\usepackage{graphicx}
\usepackage{bm}
\usepackage{xcolor}
\usepackage{float}
\usepackage{booktabs}

\usepackage{yhmath}
\usepackage{mathdots}
\usepackage{MnSymbol}
\usepackage{diagbox} 

\usepackage{pgfplots}
\pgfplotsset{compat=1.18}
\usepgfplotslibrary{fillbetween}

\everymath{\displaystyle}
\allowdisplaybreaks

\newtheorem{theorem}{Theorem}[section]

\newtheorem{proposition}{Proposition}[section]
\newtheorem{remark}{Remark}[section]

\newtheorem{corollary}{Corollary}[section]

\newcommand{\dd}{\mathrm{d}}

\providecommand{\keywords}[1]
{ \small \textbf{\textit{Keywords---}}#1 }

\title{Mathematical and Numerical Analysis of Tsunami Generation and Propagation: The Initial Tsunami Condition}
\author[$\;$]{}
\affil[$\;$]{ {\large Min CHEN}}
\affil[$\;$]{ {\scriptsize  School of Mathematical Science, Eastern Institute of Technology, Ningbo, Zhejiang 315200, China.}}
\affil[$\;$]{ {\scriptsize Email: mchen@eitech.edu.cn}}
\affil[$\;$]{ {\large }}
\affil[$\;$]{ {\large Noureddine LAMSAHEL}}
\affil[$\;$]{ {\scriptsize  School of Mathematical Science, Eastern Institute of Technology, Ningbo, Zhejiang 315200, China.}}

\affil[$\;$]{ {\scriptsize Emails: noureddine199711@gmail.com, noureddine@eitech.edu.cn }}

\date{}

\begin{document}

\maketitle

\begin{abstract}
This work investigates the appropriate initial condition for tsunami propagation that allows replacing a kinematic system with its corresponding static formulation, where the bottom is fixed at its final earthquake-induced deformation. Using the two-dimensional (2D) nonlinear shallow-water equations (NSW) and a 2D dispersive $abcd$-Boussinesq system, we prove that for the NSW system, the appropriate initial free-surface condition is exactly the final bottom deformation. For the Boussinesq model, this condition is obtained by first applying a \textit{ Boussinesq filtering operator} to the bottom profile. Furthermore, for both models, we establish an error estimate between the kinematic
and static solutions that depends linearly on the earthquake rise time. This result provides a rigorous mathematical explanation for the errors observed in previous numerical tsunami simulations. As a direct consequence, the error vanishes as the bottom motion becomes instantaneous. We validate the theoretical results using an Okada solution for the seabed deformation. In particular, both the NSW model and the passive approach, which initializes the free surface with the unfiltered bottom deformation, overestimate the amplitudes of the tsunami waves relative to the linear potential theory for the configurations considered.
\end{abstract}

\noindent\keywords{ Tsunami; Earthquake; Shallow-water model; $abcd-$Boussinesq model; Initial tsunami condition; Kinematic system; Static system,  Boussinesq filter.}

\section{Introduction}\label{intro-paper}
Tsunamis generated by submarine earthquakes are among the most disastrous natural phenomena, posing a constant threat to coastal communities \cite{synolakis2006}. In the open ocean, these waves propagate at high speeds with relatively small amplitudes; as they approach the nearshore region, however, their speed decreases while their amplitude grows significantly \cite{levin2009,saito2019tsunami,synolakis1991green}. Over recent decades, tsunami-related fatalities and economic damage have increased \cite{imamura2019}, and catastrophic events such as the 2004 Indian Ocean and  2011 Tohoku tsunamis illustrate the devastating scale of such hazards. These events have long served as a primary motivation for research in this field   \cite{kubota2026,outiskt2026,zhao2024}.

Modeling tsunamis is a challenging problem that requires computing the seafloor deformation caused by the earthquake using elastic dislocation theory \cite{dutykh2007water,okada1985surface,okada1992internal}, and  injecting the resulting bottom motion into a water wave model to predict the wave amplitude and arrival time \cite{wang2021}. Once the bottom motion is defined, selecting a suitable tsunami model requires balancing numerical feasibility with the need to accurately capture both nonlinear and dispersive effects \cite{nhess-13-1507-2013}. Although the linear and nonlinear shallow-water equations are widely used for tsunami generation and propagation \cite{dutykh2009tsunami,dutykh2011,lotto2017,MELKIOR2026114990}, it is well known that, in intermediate- to deep-water settings,  NSW systems are insufficient  owing to the dispersive effects during propagation \cite{coulaud2025comparison,nhess-13-1507-2013,saito2020tsunami,zhou2012}. We therefore also consider, in this work, a weakly dispersive Boussinesq-type model that remains applicable near the tsunami source region.

In the literature, existing studies can be classified into two main approaches. The first consists of kinematic systems, in which the bottom profile is time-dependent and usually enters the governing equations as a  forcing term. However, implementing this approach requires knowledge of the seafloor's temporal evolution, which is commonly prescribed via a rise function \eqref{bottom-type}; see, for instance, \cite{dutykh2007water,hammack1973,saito2019tsunami}. On the other hand, the static approach requires only the final, permanent seafloor deformation, generally obtained from analytical Okada solutions (see, e.g., \cite{dutykh2007water,mitsotakis2009boussinesq,triantafyllou2024strong}).  In the kinematic case, the initial free-surface elevation and velocity are commonly set to zero, and the tsunami is generated by the moving seafloor. By contrast, the static approach directly prescribes an initial free-surface condition. This simplification is possible when the initial condition is chosen appropriately and the rise time is sufficiently small \cite{chen2020mathematical,fujiwara2015shallow,iguchi2011mathematical,kajiura1963leading,saito2019tsunami,sementsov2025effect}. However, for a fixed free-surface flow model—such as NSW equations or a Boussinesq system—and assuming zero initial velocity, the initial free-surface elevation $\zeta_0$ ($\zeta_0(X):=\zeta(0,X)$)   is not uniquely  chosen in tsunami modeling. In some studies, the surface is simply identified with the seafloor deformation,
i.e., $\zeta_0 = B$ (passive generation) \cite{dutykh2006linear,dutykh2009tsunami,dutykh2013finite,kervella2007,mitsotakis2009boussinesq,ohmachi2001,saillour2022modelling}, whereas in other studies the initial condition is obtained by applying the Kajiura filter \cite{kajiura1963leading}
to the bottom deformation, which ensures consistency with the linear potential system for sudden bottom
motions  \cite{nhess-13-1507-2013,kajiura1963leading,lovholt2012,saito2019tsunami,saito2020tsunami}. Namely, 
\begin{equation}\label{intial_Euler}
    \zeta_0(X) = \frac{1}{(2\pi)^2} \int_{\mathbb{R}^2} \frac{\widehat{B}(k) \, e^{i\,k\cdot X}}{\cosh( |k| H_0)}  \dd k,\quad \forall X\in \mathbb{R}^2,
\end{equation}
where $k=(k_x,k_y)$ denotes the wave vector, $\widehat{B}$ denotes the Fourier transform of $B$, and $H_0$ is the typical constant depth.

This difference leads to the following question: \textit{for a fixed free-surface model, how should one choose the initial tsunami condition $\zeta_0$ to replace the associated kinematic system?} Most existing studies focus on  the linear potential model, for which analytical solutions can be derived for a variety of bottom deformations \cite{dutykh2006linear,dutykh2007water,hammack1973,nosov1996,saito2013,saito2019tsunami}. However, nonlinear effects have been observed to grow during propagation  \cite{dutykh2013finite, poletto2025full,ursell1953}, motivating the need for a nonlinear analysis.

In this work, we follow the approach of \cite{chen2020mathematical} and seek an initial free-surface  elevation $\zeta_0$ for which the difference between the kinematic and static formulations is controlled by the earthquake rise time. More precisely, we fix the kinematic system with zero initial conditions and seek  $\zeta_0$ for the static system, with zero initial velocity, such that the error between the two formulations is bounded explicitly by the earthquake rise time. Importantly, in the limit of instantaneous motion and under the selected  $\zeta_0$, the two systems coincide for all $t\in (0,T)$, where $T$ is the well-posedness time, as in the linear potential theory.  Our analysis focuses on the NSW equations and an $abcd-$type Boussinesq model. Nevertheless, the approach developed here has the potential to be extended to other models, such as the Peregrine system \cite{peregrine1967long}.



\subsection{Main results}

We denote the earthquake rise time by $t_\epsilon$ and the rise function by $f_\epsilon$. We assume that the seabed motion can be described as 
\begin{equation}\label{bottom-type}
    b^\epsilon(t,X)=B(X) f_\epsilon(t),\quad \forall X\in \mathbb{R}^2,\:\forall t\in \mathbb{R},
\end{equation}
where $f_\epsilon$ satisfies the hypotheses \eqref{rising-type}-\eqref{hypotheses-type}. For now, it is worth noting that the general formulation \eqref{bottom-type} encompasses the separable examples considered in \cite{dutykh2007water,hammack1973,saito2019tsunami,sementsov2025effect}. Our main contributions can then be summarized as follows:
\begin{enumerate}
    \item Under the initial tsunami condition $\zeta(0,X)=B(X)$ for the static shallow-water system, we derive an upper bound for the error between the kinematic and static systems (Theorems~\ref{convergence-NSW}--\ref{NSW_Estimate_linear}). Consequently, both systems coincide for all $t$ in $(0,T)$ when considering an instantaneous bottom motion (Corollary~\ref{NSW-equality}).
    \item We extend the Boussinesq model results from \cite{chen2020mathematical} to two-dimensional and higher-order Sobolev spaces, including local well-posedness (Theorem \ref{uniqtheorem}) and the upper bound estimates (Theorems \ref{convergence}--\ref{bous_Estimate_linear}).  Unlike in the shallow-water case, the appropriate initial condition for the static Boussinesq system is obtained by applying a \textit{Boussinesq filter} (see Remark \ref{Bous_filter_Linear_filter}) to the bottom profile.
    \item Implementing the two models based on \cite{chen2009numerical}, we perform several numerical tests using an Okada bottom profile to validate the theoretical results. These tests also highlight the impact of the passive approach on the Boussinesq model, as well as the speed of convergence of the kinematic systems towards the corresponding static systems.
\end{enumerate}

\subsection{Outline of the paper}
In Section~\ref{models_dis}, we briefly review the derivation and dispersive properties of both the NSW model and the $abcd$-Boussinesq model. Crucially, this section provides the necessary relationships between the nondimensional variables we employ throughout the paper and their corresponding physical equivalents. In Section \ref{shalow}, we delve into the analysis of the appropriate initial tsunami condition for the NSW model. The section concludes with the derivation of analytical solutions and explicit upper bounds for the linear systems. In Section \ref{bous}, our focus shifts to the considered $abcd$ model. Firstly, we establish the local well-posedness of this model; we then follow a similar analytical approach to that of Section \ref{shalow}. In Section \ref{numericalanal}, we present numerical validation results and comparisons against the linear Euler solution. Finally, Section \ref{sec:conclusions} concludes the paper by summarizing the main results, providing some additional comments, and outlining potential future work.


\section{Models of Fluid Dynamics}\label{models_dis}
This section highlights the dispersive properties of the Boussinesq model relative to the NSW model. For complete details on the derivation and validation of both models, we refer the reader to \cite{bona2002boussinesq,bona2004,chen2003equations,lannes2013water,mitsotakis2009boussinesq} and the references therein.

\begin{figure}
	\centering
	\begin{tikzpicture}
		\begin{axis}[
			axis lines = middle,
			xlabel = $X$,
			ylabel = $z$,
			xtick =  \empty,
			ytick = \empty,
			xmin = -1,
			xmax = 10,
			ymin = -3, 
			ymax = 15, 
			no markers,
			grid = none,
			every axis x label/.style={at={(current axis.right of origin)},anchor=north east},
			every axis y label/.style={at={(current axis.above origin)},anchor=north east},
			]
            \pgfmathdeclarefunction{b}{1}{%
			\pgfmathparse{ (x>4 && x<7.1) ? cos(deg(x)*2)+2 : 2 }%
		}
			
			\addplot[black, domain=-1:10, samples=100] {sin(deg(x))*cos(deg(x*2))+10} node[pos=0.7, anchor=south west] {$\zeta(t,X) $};
			\addplot[black, domain=-1:9, samples=100, name path=A] {0} ;
			
			\addplot[dashed, domain=-1:9, samples=100, name path=B]{ b(x) }  node[pos=0.7, anchor=south west] {$-H_0+b(t,X) $}; 
			\draw[dashed] (axis cs:-1,10) -- (axis cs:10,10)  node[pos=0.05, above] {0};
		
			\addplot[gray!20] fill between[of=A and B];
			
		\end{axis}
	\end{tikzpicture}
	\caption{{\footnotesize Scheme for the domain.}}
	\label{fig:myplot1}
\end{figure}
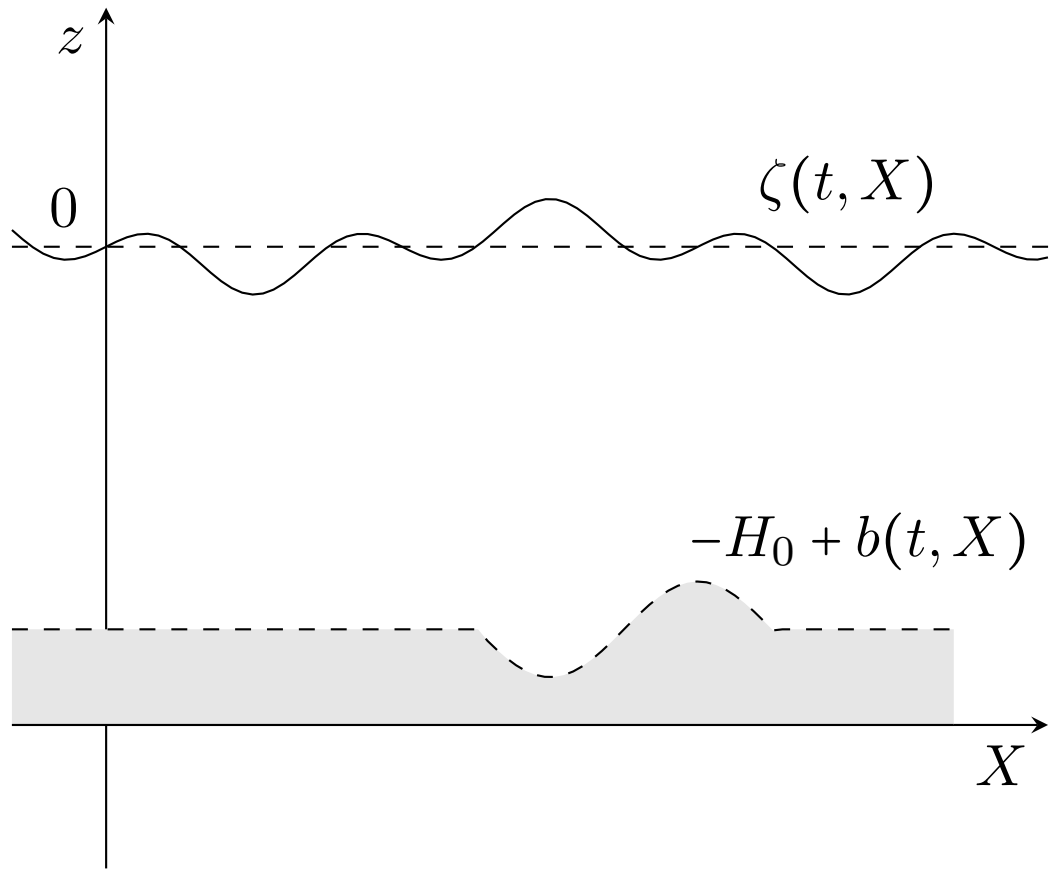

We assume that tsunami motion is governed by the free-surface Euler equations for an inviscid, incompressible, and irrotational fluid \cite{lannes2013water}. The fluid domain is bounded below by a moving bottom and above by a free surface, with both boundaries described by graphs of functions; see Figure \ref{fig:myplot1}.

We denote by $A_0$  and $\lambda_0$  the typical wave amplitude and wavelength, respectively, and introduce the following nondimensional parameters:

$$\varepsilon=\frac{A_0}{H_0},\quad \sigma=\frac{H_0^2}{\lambda_0^2}.$$
Here, $\varepsilon$  characterizes the nonlinearity and usually arises in the nonlinear terms of the Boussinesq equations; $\sigma$ characterizes the dispersion and appears in the dispersive terms. To illustrate the influence of these variables on the Euler system \cite{lannes2013water}, we introduce the following nondimensional scaled variables; see, e.g., \cite{chen2003equations}
\begin{equation}\label{nondimensional_1}
    \widetilde{X}:=\left(\widetilde{x},\widetilde{y}\right)=\frac{1}{\lambda_0}X,\quad\; \widetilde{z}=\frac{z}{H_0},\quad \widetilde{t}=\frac{\sqrt{g\,H_0}}{\lambda_0}t,
\end{equation}
and 
\begin{equation}\label{nondimensional_2}
   (\widetilde{u},\widetilde{v})=\frac{H_0}{A_0\sqrt{g\,H_0}}(u,v),\quad \widetilde{w}=\frac{\lambda_0}{A_0\sqrt{g\,H_0}}w,\quad\widetilde{\zeta}=\frac{\zeta}{A_0},\quad \widetilde{b}=\frac{b}{A_0}.
\end{equation}
By introducing these changes of variables into the Euler equations and averaging over the vertical direction, we obtain the nondispersive $\mathcal{O}(\sigma)$ shallow-water equations in nondimensional form \cite{chen2003equations,lannes2013water}:
\begin{equation}\label{pd1}
	\left\{
	\begin{array}{ll}
		\widetilde{\zeta}_{\widetilde{t}}+\widetilde{\nabla}\cdot\left(\widetilde{H}\,\overline{U}\right)= \widetilde{b}_{\widetilde{t}},\vspace{0.3cm}\\
        \overline{U}_{\widetilde{t}}+\widetilde{\nabla} \widetilde{\zeta} +\varepsilon(\overline{U}.\widetilde{\nabla})\overline{U}=\mathbf{0},
	\end{array}
	\right.
\end{equation}
 where $\widetilde{H}=1+\varepsilon(\widetilde{\zeta}-\widetilde{b})$ represents the nondimensional fluid depth and $\overline{U}$ denotes the depth-averaged horizontal velocity given by 
$$\overline{U}=\frac{1}{\widetilde{H}}\displaystyle\int_{-1+\varepsilon \widetilde{b}}^{\varepsilon \widetilde{\zeta}}(\widetilde{u},\widetilde{v})\,\dd \widetilde{z}.$$

The above model \eqref{pd1} can be regarded as the simplest fully nonlinear approximation of the full Euler system, providing a powerful tool for the analysis of waves in the near-shore region, including tsunami run-up \cite{grilli1994shoaling,synolakis1991green}. However, this model neglects dispersive effects, which may impact tsunami analysis, as discussed in  Section \ref{intro-paper}.

Let us assume a long-wave and small-amplitude regime, which in particular includes tsunamis \cite{saito2019tsunami}. Namely
$$ \varepsilon \ll 1,\quad \sigma \ll 1,\quad \text{and}\quad  \varepsilon=\mathcal{O}(\sigma).$$
By similar arguments used to derive system \eqref{pd1} from Euler equations, however, this time we keep some dispersion and only neglect the terms of order $\mathcal{O}(\varepsilon\sigma,\sigma^2)$, we obtain the following; see \cite{chen2003equations,mitsotakis2009boussinesq}:
\begin{equation}\label{pd2}
	\left\{
	\begin{array}{ll}
		\widetilde{\zeta}_{\widetilde{t}}+\widetilde{\nabla}\cdot\left(\widetilde{H}\,\widetilde{U}\right)-\frac{\sigma}{3}\widetilde{\Delta} \widetilde{\zeta}_{\widetilde{t}}-\frac{\sigma}{6}\;\widetilde{\Delta} \widetilde{b}_{\widetilde{t}}=\widetilde{b}_{\widetilde{t}},\vspace{0.3cm}\\
        \widetilde{U}_{\widetilde{t}}+\widetilde{\nabla} \widetilde{\zeta} +\varepsilon(\widetilde{U}.\widetilde{\nabla})\widetilde{U}=\mathbf{0},
	\end{array}
	\right.
\end{equation}
where $\widetilde{U}$ is the horizontal velocity at the free surface $\widetilde{\zeta}$. The above system \eqref{pd2} is a particular case of the general $abcd$ model \cite{chen2003equations}, where $\theta=1$ is chosen as in \cite{chen2020mathematical} in order to prevent the emergence of a velocity singularity in the subsequent analysis; the general case $\theta\neq 1$ will be considered in future work; see Section \ref{sec:conclusions}.

Before presenting the dispersive comparison, it is important to make the following:
\begin{remark}
    \begin{itemize}
        \item Systems \eqref{pd1} and \eqref{pd2}  are linked through the relation $\widetilde{U}=\overline{U}+\mathcal{O}(\sigma)$. Substituting this relation into \eqref{pd2} and neglecting the dispersive terms of order $\mathcal{O}(\sigma)$ yields the system \eqref{pd1}.
        \item Since system \eqref{pd2} is characterized at the free surface, it is particularly well suited to the inverse problem of underwater bottom detection from surface measurements, which require the surface velocity; see, e.g., \cite{fontelos2017bottom,lamsahel2026uniqueness,lecaros2024optimal}.
    \end{itemize}
\end{remark}

\begin{figure}[!htb]
  \centering
  \includegraphics[width=0.7\textwidth]{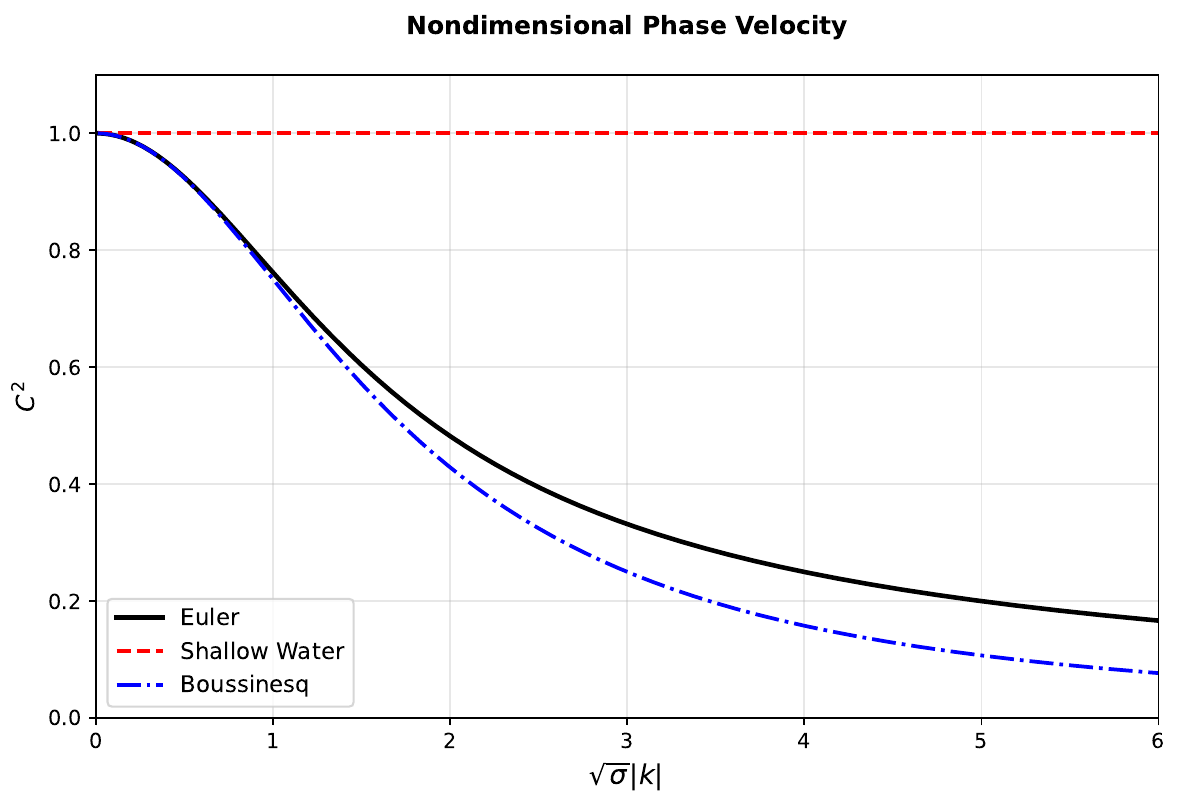}
  \caption{ {\footnotesize Comparison of linear dispersion relations.}}
  \label{fig: 1}
\end{figure}

We now present a comparison of the dispersion relations of the
Euler equations \cite{lannes2013water}, \eqref{pd1}, and \eqref{pd2}. Assuming a constant bottom profile $b=0$, $\varepsilon=0$, and considering plane wave solutions, we obtain the following dispersion relations; see \cite{bona2002boussinesq,coulaud2025comparison} for details.

\begin{equation}\label{pd3}
\left\{
\begin{array}{l}
\displaystyle
C^{2}_{\rm Euler}
=\frac{1}{k\sqrt{\sigma}}\tanh\!\bigl(k\sqrt{\sigma}\,\bigr)
=1-\frac{\sigma k^{2}}{3}+\frac{2\sigma^{2}k^{4}}{15}
+\mathcal{O}\!\bigl(\sigma^{3}k^{6}\bigr),\vspace{0.2cm}\\
\displaystyle
C^{2}_{\rm NSW}=1,\\
\displaystyle
C^{2}_{\rm Boussinesq}
=\frac{1}{1+\dfrac{\sigma k^{2}}{3}}
=1-\frac{\sigma k^{2}}{3}+\frac{\sigma^{2}k^{4}}{9}
+\mathcal{O}\!\bigl(\sigma^{3}k^{6}\bigr).
\end{array}
\right.
\end{equation}
Clearly, from Figure \ref{fig: 1} and from  the third equation in \eqref{pd3}, the Boussinesq system \eqref{pd2} is a second-order dispersive model, similar to the Peregrine model \cite{peregrine1967long}, which was used in \cite{saito2010,saito2020tsunami} for tsunami propagation, whereas the shallow-water model is non-dispersive. Further comparisons with other Boussinesq models can be found in \cite{coulaud2025comparison}.

We conclude this section with the following crucial remark.
\begin{remark}\label{nondimensional_change}
   Since tsunami generation and propagation are typical long-wave small-amplitude phenomena, in this paper we adopt the following 'standard' change of variables: 
   \begin{equation}\label{nondimensional_3}
  \overbrace{X}=\sigma^{-\frac{1}{2}} \widetilde{X},\quad\overbrace{t}=\sigma^{-\frac{1}{2}} \widetilde{t},\quad (\overbrace{u},\overbrace{v})=\varepsilon(\widetilde{u},\widetilde{v}),\quad \overbrace{\zeta}=\varepsilon \widetilde{\zeta},\quad \overbrace{b}=\varepsilon\widetilde{b}.
\end{equation}
We refer the reader to \cite{bona2002boussinesq,chen2003equations} for further details. In particular, the relationship between the physical bottom $b$ and the new nondimensional bottom is given by
\begin{equation}\label{relation_bottom_phyiscal_bottom}
    \overbrace{b}(\overbrace{t},\overbrace{X})=\frac{1}{H_0} b\left(\frac{H_0}{\sqrt{gH_0}}\overbrace{t},H_0\overbrace{X}\right).
\end{equation}
For the sake of clarity, we henceforth omit the overbrace notation $\overbrace{\square}$.
\end{remark}
Having provided the necessary comments on the considered models, we now proceed with the rigorous mathematical analysis of the appropriate choice of initial surface condition for tsunami propagation.

\section{The initial condition for the shallow-water model}\label{shalow}
In this section, we present the first theorem of this paper (Theorem \ref{convergence-NSW}), which provides an upper bound on the difference between the solutions of the kinematic system \eqref{pd1-S3} and the static system \eqref{pd2-S3}. As a consequence, we prove that for sudden bottom deformation, the solutions of \eqref{pd1-S3} and \eqref{pd2-S3} coincide for $t>0$  (Corollary \ref{NSW-equality}). In particular, this section provides an error estimate between the two shallow-water reformulations derived in \cite{fujiwara2015shallow,iguchi2011mathematical}.

We consider seabed deformations of the form \eqref{bottom-type}, with $f_\epsilon$  given by
\begin{equation}\label{rising-type}
 f_\epsilon(t)=   \left\{
	\begin{array}{ll}
		0,\quad t\leq 0,\\
        f_{t_\epsilon}(t),\quad 0<t\leq t_\epsilon,\\
        1,\quad t>t_\epsilon.
	\end{array}
	\right.
\end{equation}
Here, $t_\epsilon$ represents the nondimensional earthquake rise time (see Remark \ref{nondimensional_change}), and the rise function $f_{t_\epsilon}$
is assumed to satisfy the following hypotheses
\begin{equation}\label{hypotheses-type}
   \left\{
	\begin{array}{ll}
		f_{t_\epsilon}(0)=0,\; f_{t_\epsilon}(t_\epsilon)=1,\\
        f_{t_\epsilon}\in  W^{1,\infty}(0,t_\epsilon),\; f_{t_\epsilon}'\geq 0,
	\end{array}
	\right.
\end{equation}
where $W^{1,\infty}$ denotes the space of Lipschitz continuous functions.

Under some technical assumptions, the kinematic tsunami system approximating the Euler equations for a relatively slow-moving bottom  \cite{fujiwara2015shallow,iguchi2011mathematical} is given by
\begin{equation}\label{pd1-S3}
	\left\{
	\begin{array}{ll}
		\zeta^\epsilon_t+\nabla\cdot\left(H^\epsilon\,\overline{U^\epsilon}\right)= b_t^\epsilon,\vspace{0.3cm}\\
       \overline{U^\epsilon}_t+\nabla \zeta^\epsilon +(\overline{U^\epsilon}.\nabla)\overline{U^\epsilon}=\mathbf{0},\vspace{0.3cm}\\
       \zeta^\epsilon(0,X)=0,\quad \overline{U^\epsilon}(0,X)=\mathbf{0}.
	\end{array}
	\right.
\end{equation}
The static system  traditionally used in numerical tsunami analysis to replace \eqref{pd1-S3} and that approximates the Euler equations for a relatively fast-moving bottom  \cite{fujiwara2015shallow,iguchi2011mathematical} is given by
\begin{equation}\label{pd2-S3}
	\left\{
	\begin{array}{ll}
		\zeta_t+\nabla\cdot\left(H\,\overline{U}\right)= 0,\vspace{0.3cm}\\
       \overline{U}_t+\nabla \zeta +(\overline{U}.\nabla)\overline{U}=\mathbf{0},\vspace{0.3cm}\\
       \zeta(0,X)=B,\quad \overline{U}(0,X)=\mathbf{0}.
	\end{array}
	\right.
\end{equation}
In the above system, the bottom is assumed to correspond to the final earthquake deformation at \(t = t_\epsilon\), and \(H = 1 + \zeta - B\).

Since we aim to bridge the gap between theoretical and numerical analyses, we will derive an upper bound for the following relative error \eqref{relativeerror}, thereby providing a rigorous estimate of the errors observed in some numerical tsunami simulations; see, e.g., \cite{dutykh2009tsunami,kervella2007}.

\begin{equation}\label{relativeerror}
    E_s(t)=\frac{\left\Vert\zeta^\epsilon(t) -\zeta(t)\right\Vert_{H^{s-1}(\mathbb{R}^2)} +\left\Vert\overline{U^\epsilon}(t)-\overline{U}(t)\right\Vert_{H^{s-1}(\mathbb{R}^2;\mathbb{R}^2)}  }{\left\Vert  B\right\Vert_{H^{s-1}(\mathbb{R}^2)} }, \quad s\geq 1,\quad t\geq0,
\end{equation}
where $B\neq 0$. Before addressing an upper bound on the aforementioned error \eqref{relativeerror}, we require the following local well-posedness result.

\begin{proposition}\label{well-posdenss-shalow}
Let \(s>2\) and assume that \(B\in H^{s+1}(\mathbb{R}^2)\), and let \(f_\epsilon\) defined by \eqref{rising-type} satisfy \eqref{hypotheses-type}. Then there exists a time $T:=T(B,s)>0$, depending only on  $B$ and $s$, such that each of the systems \eqref{pd1-S3} and \eqref{pd2-S3} admits a unique solution $(\underline{\zeta},\underline{U})$ satisfying 
$$(\underline{\zeta},\underline{U})\in C([0,T]; H^s(\mathbb{R}^2)\times H^s(\mathbb{R}^2;\mathbb{R}^2)).$$
Moreover, there exists $R:=R(B,s)>  \left\Vert B\right\Vert_{H^{s+1}(\mathbb{R}^2)}$ , depending only on  $B$ and $s$, such that
\begin{equation}\label{enerrgy-estimate}
    \left\Vert (\underline{\zeta},\underline{U})\right\Vert_{ L^{\infty}(0,T; H^s(\mathbb{R}^2)\times H^s(\mathbb{R}^2;\mathbb{R}^2))}\leq R \quad\text{and}\quad \inf_{(0,T)\times\mathbb{R}^2} \underline{H}\geq \frac{1}{2},
\end{equation}
where $\underline{\square}$ refers to \eqref{pd1-S3} or \eqref{pd2-S3}.

\end{proposition}

\begin{proof}
    Systems \eqref{pd1-S3} and \eqref{pd2-S3} belong to the so-called symmetrized hyperbolic systems \cite{taylor1997nonlinear}. The proof of Proposition \ref{well-posdenss-shalow} for \eqref{pd2-S3} can be found in \cite[Proposition 6.1]{lannes2013water}. For system \eqref{pd1-S3}, we introduce the change of variables $\eta^\epsilon=\zeta^\epsilon-b^\epsilon$; then \eqref{pd1-S3} is equivalent to
    \begin{equation}\label{pd3-S3}
	\left\{
	\begin{array}{ll}
		\eta^\epsilon_t+\nabla\cdot\left(H^\epsilon\,\overline{U^\epsilon}\right)= 0,\vspace{0.3cm}\\
       \overline{U^\epsilon}_t+\nabla \eta^\epsilon +(\overline{U^\epsilon}.\nabla)\overline{U^\epsilon}=-\nabla b^\epsilon,\vspace{0.3cm}\\
       \eta^\epsilon(0,X)=0,\quad \overline{U^\epsilon}(0,X)=\mathbf{0},
	\end{array}
	\right.
\end{equation}
where $H^\epsilon=1+\eta^\epsilon$ with $H^\epsilon(0,\cdot)=1$. Importantly, the new forcing term involves \(f_\epsilon\) rather than \(f_\epsilon'\). Following arguments similar to those in \cite[Chapter 6]{lannes2013water}, and taking into account the forcing term \((0,-f_\epsilon\nabla B)^T\), together with \(\|f_\epsilon\|_{L^\infty} \leq 1\) and zero initial conditions, we apply the symmetrizer
\begin{equation}\label{symmertrizer}
    S\left(\eta^\epsilon\right)=\begin{pmatrix}
        1 & 0 & 0\\
        0 & H^\epsilon & 0\\
        0 & 0 & H^\epsilon
    \end{pmatrix},
\end{equation}
and obtain the desired result. The details are omitted; we refer the reader to \cite[Section 6.1.1]{lannes2013water}.
\end{proof}

We are now in a position to state the first main result of the present work.
\begin{theorem}\label{convergence-NSW}
    Let $s>2$ and assume that the hypotheses of Proposition \ref{well-posdenss-shalow} are satisfied. Then there exists a constant $C_{Sw}$ depending only on $T$, $R$, $B$,  and  $s$, such that the relative error defined by \eqref{relativeerror} satisfies
    \begin{equation}\label{upper-bound-error-NSW}
        E_s(t)\leq ( 1-f_\epsilon(t))+ C_{Sw}\left\Vert 1-f_\epsilon\right\Vert_{L^1(0,t) }  ,\quad \forall t\in (0,T),
    \end{equation}
    where $f_\epsilon$ is given by \eqref{rising-type}. In particular, for all $t\in ( t_\epsilon,T)$, where $t_\epsilon$ denotes the earthquake rise time  \eqref{rising-type}, we have

    \begin{equation}\label{after-ert-upper-bound-error-NSW}
        E_{s_1}(t)\leq C_{Sw}\, t_\epsilon,\quad  \forall s_1\in [1,s].
    \end{equation}
\end{theorem}
\begin{proof}
Fix $s>2$, and let $V^\epsilon=(\eta^\epsilon,\overline{U^\epsilon})^T$  denote the unique solution of \eqref{pd3-S3} given by Proposition \ref{well-posdenss-shalow}. Recall that $\eta^\epsilon=\zeta^\epsilon-b^\epsilon$, where $\zeta^\epsilon$ is the solution of \eqref{pd1-S3}.\\
Multiplying \eqref{pd3-S3} by the symmetrizer $S$ given in \eqref{symmertrizer} and performing straightforward computations yields the following equivalent system:\begin{equation}\label{h-kinematic-NSW}
    S(\eta^\epsilon)\partial_t V^\epsilon+A_1(V^\epsilon)\partial_1 V^\epsilon+A_2(V^\epsilon)\partial_2 V^\epsilon= (0,- H^\epsilon\nabla b^\epsilon)^T,
\end{equation}
where $\partial_1:=\partial_x$, $\partial_2:=\partial_y$ and 
\begin{equation}\label{matrices-hypo}
   A_1(V^\epsilon) =\begin{pmatrix}
\, \overline{u_1^\epsilon}& \,H^\epsilon&0\\
 \,H^\epsilon & H^\epsilon \,\overline{u_1^\epsilon} &0\\
0&0&  H^\epsilon \,\overline{u_1^\epsilon}
\end{pmatrix}, \;\;   A_2(V^\epsilon) =\begin{pmatrix}
\,  \overline{u_2^\epsilon}&0& \,H^\epsilon\\
0 &  H^\epsilon \,\overline{u_2^\epsilon} &0\\
 \,H^\epsilon&0&  H^\epsilon \,\overline{u_2^\epsilon}
\end{pmatrix}
\quad \text{and} \quad \overline{U^\epsilon}=(\overline{u_1^\epsilon},\overline{u_2^\epsilon})^T.
\end{equation}
Setting \(\eta=\zeta-B\) and following the same arguments as above, we obtain the following equivalent symmetric hyperbolic reformulation of \eqref{pd2-S3}.
\begin{equation}\label{h-static-NSW}
    S(\eta)\partial_t V+A_1(V)\partial_1 V+A_2(V)\partial_2 V= (0,- H\nabla B)^T,
\end{equation}
We now define the error \(E=V^\epsilon-V\). Subtracting the two equations \eqref{h-kinematic-NSW} and \eqref{h-static-NSW}, we obtain the following:
$$S(\eta^\epsilon)\partial_t E+A_1(V^\epsilon)\partial_1 E+A_2(V^\epsilon)\partial_2 E=F,$$
where 
$$F=F_1-\left(S(\eta^\epsilon)-S(\eta)\right)\partial_t V-\displaystyle\sum_{j=1,2}\left(A_j (V^\epsilon)-A_j (V)\right)\partial_j V,$$
and 
$$F_1=-\left(0, \nabla B\left\{H^\epsilon(f_\epsilon-1)+\eta^\epsilon-\eta\right\}\right)^T.$$
Using arguments similar to those in \cite[Section 6.1.1]{lannes2013water}, together with the fact that \(E(0,X)=\mathbf{0},\;\forall X\in \mathbb{R}^2\), we derive the following energy estimate:
\begin{equation}\label{energy-estima}
    \left\Vert E(t)\right\Vert_{H^{s-1}}\leq  C \displaystyle \int_0^t \left(\left\Vert F_1\right\Vert_{H^{s-1}}+ \left\Vert E\right\Vert_{H^{s-1}}\right)(\tau)\; \dd\tau,
\end{equation}
where \(C\) depends only on \(T\), \(R\), and \(s\). Here and in the subsequent steps of the proof, we adopt the following notation:
\begin{equation}\label{notation}
\begin{gathered}
  \text{The symbol } H^s \text{ denotes either }
  H^s(\mathbb{R}^2) \text{ or } H^s(\mathbb{R}^2;\mathbb{R}^2),\\
  \text{depending on the context. Moreover, we adopt the product-sum norm.}
\end{gathered}
\end{equation}
We can derive the following upper bound for $F_1$:
\begin{equation}\label{observation-NSW}
    \left\Vert F_1\right\Vert_{H^{s-1}}\leq    \left\Vert \nabla B\right\Vert_{H^{s-1}}\left((1+2\, C_s R) (1-f_\epsilon) + C_s\left\Vert E\right\Vert_{H^{s-1}}\right).
\end{equation}
Substituting the above inequality \eqref{observation-NSW} into estimate \eqref{energy-estima} and using the fact that $R>\left\Vert B\right\Vert_{H^{s+1}}$, we obtain
$$ \left\Vert E(t)\right\Vert_{H^{s-1}}\leq C_1 \left\Vert \nabla B\right\Vert_{H^{s-1}}\displaystyle\int_0^t (1-f_\epsilon(\tau))\;\dd \tau\;+C_2 \displaystyle\int_0^t \left\Vert E(\tau)\right\Vert_{H^{s-1}}\;\dd \tau,$$
where $C_1=C (1+2C_s R)$ and $C_2= C(1+C_sR)$ .\\
An application of Gronwall's inequality yields
\begin{equation}\label{Granweel-est}
    \left\Vert E(t)\right\Vert_{H^{s-1}}\leq C_1 e^{C_2 T}\,\left\Vert \nabla B\right\Vert_{H^{s-1}}\displaystyle\int_0^t (1-f_\epsilon(\tau))\;\dd\tau,\quad \forall t\in (0,T). 
\end{equation}
We now observe that the relative error defined in \eqref{relativeerror} satisfies
\begin{equation}\label{inter}
   E_s(t)\leq \frac{\left\Vert E(t)\right\Vert_{H^{s-1}}}{\left\Vert  B\right\Vert_{H^{s-1}}}+ (1-f_\epsilon(t)), \quad\forall t\in (0,T). 
\end{equation}
Substituting inequality \eqref{Granweel-est} into the right-hand side of the above relation \eqref{inter}, we obtain the desired estimate \eqref{upper-bound-error-NSW}. Moreover, we combine the Sobolev embedding with the fact that, for \(t\geq t_\epsilon\), we have
$$f_\epsilon(t)=1\quad\text{and}\quad \left\Vert 1-f_\epsilon\right\Vert_{L^1(0,t)}\leq t_\epsilon.$$
This yields \eqref{after-ert-upper-bound-error-NSW} and completes the proof.
\end{proof}
Before addressing the linear case (nonlinearity $\varepsilon\approx0$), we first present the result concerning the instantaneous motion of the bottom. To this end, we consider a bottom profile of the form
\begin{equation}\label{insta_bottom}
b^0(t,X)=\chi(t)B(X),\quad \forall t\in  \mathbb{R} ,\;\forall X\in \mathbb{R}^2,
\end{equation}
where \(\chi(t)\) denotes the Heaviside function. Then, the instantaneous kinematic system reads
\begin{equation}\label{pd1-S3_inst}
	\left\{
	\begin{array}{ll}
		\zeta^0_t+\nabla\cdot\left(H^0\,\overline{U^0}\right)= \delta_0 B,\vspace{0.3cm}\\
       \overline{U^0}_t+\nabla \zeta^0 +(\overline{U^0}.\nabla)\overline{U^0}=\mathbf{0},\vspace{0.3cm}\\
       \zeta^0(0,X)=0,\quad \overline{U^0}(0,X)=\mathbf{0},
	\end{array}
	\right.
\end{equation}
where $H^0=1+\zeta^0-b^0$, $\delta_0$  is the Dirac mass, $t=0$ represents the impulse state, and the solution $(\zeta^0,\overline{U^0})$ vanishes for all $t\leq 0$. Then we have 

\begin{corollary}\label{NSW-equality}
Let $s>2$ and assume that the hypotheses of Proposition \ref{well-posdenss-shalow} are satisfied. Then the unique solution of \eqref{pd1-S3_inst} is given by
\[
(\zeta^0(t),\overline{U}^0(t))
=
\begin{cases}
(0,0), & t\leq 0,\\[1mm]
(\zeta(t),\overline{U}(t)), & 0<t\leq T,
\end{cases}
\]
where $(\zeta,\overline{U})$ denotes the solution of the static
system \eqref{pd2-S3}. In particular,
\[
\lim_{t\to 0^+}
\bigl(\zeta^0(t),\overline{U}^0(t)\bigr)
=
(B,0).
\]
Moreover, for every $\delta\in(0,T)$, we have
\[
(\zeta^\epsilon,\overline{U}^\epsilon)
\xrightarrow[t_\epsilon\to 0]{}
(\zeta^0,\overline{U}^0)
\quad\text{in}\quad
C\!\left(
[\delta,T];
H^{s-1}(\mathbb{R}^2)
\times
H^{s-1}(\mathbb{R}^2;\mathbb{R}^2)
\right),
\]
where $(\zeta^\epsilon,\overline{U}^\epsilon)$ denotes the
solution of the (regularized) kinematic system
\eqref{pd1-S3}.
\end{corollary}
\begin{proof}
The result is a direct consequence of Proposition~\ref{well-posdenss-shalow} and Theorem~\ref{convergence-NSW}. The remaining standard details are omitted for brevity.
\end{proof}

 In tsunami applications, the nonlinearity parameter $\varepsilon$ is typically small, especially near the source region, as the water depth ranges from approximately 1 km to 4 km, whereas the wave amplitude is generally of the order of 1 m. For instance, in the case of the 2004 Indian Ocean tsunami, $\varepsilon$ is of order $10^{-4}$, see \cite[Example 1.5]{ lannes2013water}. Although weak nonlinear effects can accumulate over propagation time (see, e.g., \cite{dutykh2013finite,poletto2025full}), we perform a linear analysis, which yields the analytical solutions and consequently  sharper upper bounds.

Neglecting the nonlinear terms, systems \eqref{pd1-S3} and \eqref{pd2-S3} reduce respectively to the following wave systems
\begin{equation}\label{pd1-S3-linear}
	\left\{
	\begin{array}{ll}
		\zeta^\epsilon_t+\nabla\cdot\overline{U^\epsilon}= b_t^\epsilon,\vspace{0.3cm}\\
       \overline{U^\epsilon}_t+\nabla \zeta^\epsilon =\mathbf{0},\vspace{0.3cm}\\
       \zeta^\epsilon(0,X)=0,\quad \overline{U^\epsilon}(0,X)=\mathbf{0},
	\end{array}
	\right.
\end{equation}
and 
\begin{equation}\label{pd2-S3-linear}
	\left\{
	\begin{array}{ll}
		\zeta_t+\nabla\cdot\overline{U}= 0,\vspace{0.3cm}\\
       \overline{U}_t+\nabla \zeta =\mathbf{0},\vspace{0.3cm}\\
       \zeta(0,X)=B(X),\quad \overline{U}(0,X)=\mathbf{0}.
	\end{array}
	\right.
\end{equation}
By exploiting the fact that the above systems can be solved analytically, we obtain the following result.
\begin{theorem}\label{NSW_Estimate_linear}
   Let $s\geq 0$ and $T>0$, and assume that $B\in H^{s+1}(\mathbb{R}^2)$. Then, for every $t\in(0,T)$ and $k\in \mathbb{R}^2$, the solutions of systems \eqref{pd1-S3-linear} and \eqref{pd2-S3-linear} are respectively given by
   \begin{equation}\label{pd1-S3-linear_solutions}
	\left\{
	\begin{array}{ll}
		\widehat{\zeta^\epsilon}(t,k)=\left\{\displaystyle\int_0^t \cos\left(|k| (t-s)\right)f'_\epsilon(s)\;\dd s\right\}\widehat{B}(k),\quad \widehat{\overline{U^\epsilon}}(t,k)=-i\left\{\displaystyle\int_0^t \widehat{\zeta^\epsilon}(s,k) \,\dd s\right\} k,\vspace{0.3cm}\\
       \widehat{\zeta}(t,k)= \cos\left(|k|t\right)\widehat{B}(k),\quad \widehat{\overline{U}}(t,k)=-i\left\{\displaystyle\int_0^t \widehat{\zeta}(s,k) \,\dd s\right\} k. \\
       
	\end{array}
	\right.
\end{equation}
 Furthermore, for every $t\in(t_\epsilon,T)$, the following estimates hold:
\begin{equation}\label{estimate_linear}
    \left\Vert\zeta^\epsilon(t) -\zeta(t)\right\Vert_{H^{s}(\mathbb{R}^2)} \leq \left\Vert  \nabla B\right\Vert_{H^{s}(\mathbb{R}^2)}\;t_\epsilon
\end{equation}
and 
\begin{equation}\label{estimate_linear_velocity}
    \left\Vert\overline{U^\epsilon}(t) -\overline{U}(t)\right\Vert_{H^{s}(\mathbb{R}^2)} \leq \left\Vert \nabla B\right\Vert_{H^{s}(\mathbb{R}^2)} \,t_\epsilon
\end{equation}
\end{theorem}
\begin{proof}
    Applying the Fourier transform to system \eqref{pd1-S3-linear}, we obtain
    \begin{equation*}
	\left\{
	\begin{array}{ll}
		\widehat{\zeta^\epsilon}_t+ik\cdot\widehat{\overline{U^\epsilon}}= f_\epsilon' \widehat{B},\vspace{0.3cm}\\
       \widehat{\overline{U^\epsilon}}_t+i\widehat{\zeta^\epsilon}k =\mathbf{0},\vspace{0.3cm}\\
       \widehat{\zeta^\epsilon}(0,k)=0,\quad \widehat{\overline{U^\epsilon}}(0,k)=\mathbf{0}.
	\end{array}
	\right.
\end{equation*}
The above system is equivalent to

\[
 \begin{pmatrix}
\widehat{\zeta^{\epsilon}} \\[2pt]
\widehat{\overline{U^{\epsilon}}}
\end{pmatrix}_t
=
\begin{pmatrix}
0 & -i k \\
-i k^T & \mathbf{0}_{2\times 2}
\end{pmatrix}\begin{pmatrix}
\widehat{\zeta^{\epsilon}} \\[2pt]
\widehat{\overline{U^{\epsilon}}}
\end{pmatrix}
+
f_{\epsilon}'(t)
\begin{pmatrix}
\widehat{B}(k) \\
\mathbf{0}
\end{pmatrix},\quad \begin{pmatrix}
\widehat{\zeta^{\epsilon}} \\[2pt]
\widehat{\overline{U^{\epsilon}}}
\end{pmatrix}(0,k)=\mathbf{0}.
\]
Using Duhamel's formula on the above ODE yields the explicit solution \eqref{pd1-S3-linear_solutions}. More precisely, we obtain the following equivalent expression
for the velocity:
\begin{equation}\label{proof_linear1}
\widehat{\overline{U^\epsilon}}(t,k)=-i\frac{k}{|k|}\left\{\displaystyle\int_0^t \sin\left(|k| (t-s)\right)f'_\epsilon(s)\;\dd s\right\}\widehat{B}(k),\quad k\neq 0.
\end{equation}

Applying the same arguments to system \eqref{pd2-S3-linear} yields its explicit solution, with 
\begin{equation}\label{proof_linear2}
    \widehat{\overline{U}}(t,k)= -i\frac{k}{|k|}\sin\left(|k|t\right)\widehat{B}(k),\quad k\neq 0.
\end{equation}
We now establish the estimate \eqref{estimate_linear}. Let $t\in (0,T)$ and $k\in \mathbb{R}^2$. Subtracting the solutions of \eqref{pd1-S3-linear} and \eqref{pd2-S3-linear} yields
\begin{equation}\label{estimat_proof}
     \left|\widehat{\zeta^\epsilon}(t,k)- \widehat{\zeta}(t,k)\right|\leq \left|\widehat{B}(k)\right|\displaystyle\int_0^t \left| \cos\left(|k| (t-s)\right)- \cos\left(|k| t\right)\right| f'_\epsilon(s) \;\dd s+ (1-f_\epsilon(t)) \left|\widehat{B}(k)\right|
    \end{equation}
By the mean value theorem, we have
$$\left| \cos\left(|k| (t-s)\right)- \cos\left(|k| t\right)\right|\leq |k| s,\quad \forall s\in (0,t).$$
Injecting the above estimate into \eqref{estimat_proof}, we obtain
\begin{equation}\label{importnatstep}
    \left|\widehat{\zeta^\epsilon}(t,k)- \widehat{\zeta}(t,k)\right|\leq  \left\{t_\epsilon-\displaystyle\int_0^{t_\epsilon} f_\epsilon (s)\,\dd s\right\}\left|k\right|\left|\widehat{B}(k)\right|+ (1-f_\epsilon(t)) \left|\widehat{B}(k)\right|.
\end{equation}
Using the fact that $f_\epsilon=1$ for $t>t_\epsilon$, we obtain \eqref{estimate_linear}. \\
Following similar arguments and using the velocity expressions
\eqref{proof_linear1} and \eqref{proof_linear2}, we obtain the upper bound \eqref{estimate_linear_velocity}. This completes the proof.
\end{proof}
We conclude this section with a remark concerning the
linear analysis.
\begin{remark}
The assumption $B\in H^{s+1}(\mathbb{R}^2)$ in the above Theorem \ref{NSW_Estimate_linear} can be relaxed
to $B\in H^s(\mathbb{R}^2)$ for $s\geq 0$ through a density
argument, as the error estimates (see, e.g., \eqref{estimat_proof}) are controlled by the
$H^s$-norm of $B$; see \cite[proof of Theorem 2.1]{chen2020mathematical}
for details in one dimension. However, this relaxation introduces an additional small
$H^s$-term in the upper bound that does not depend
on $t_\epsilon$.
\end{remark}
Having rigorously quantified the error between the kinematic and static shallow-water systems, we now turn to the model \eqref{pd2}.


\section{The initial condition for the Boussinesq model}\label{bous}
One of the principal objectives of this section is to extend and improve the results of \cite[Section 4]{chen2020mathematical}. We begin by establishing the local well-posedness of systems \eqref{Bous-Pd1} and \eqref{Bous-Pd2}; see Theorem \ref{uniqtheorem}. We then derive an upper bound for the relative error \(E_s^{Bous}\) defined in \eqref{Error-bous}; see Theorem \ref{convergence}. 

 In the context of tsunami generation and propagation \cite{chen2020mathematical,mitsotakis2009boussinesq}, the kinematic system associated with \eqref{pd2}, after the change of variables in Remark \ref{nondimensional_change}, reads
\begin{equation}\label{Bous-Pd1}
	\left\{
	\begin{array}{ll}
		\zeta_t^\epsilon+\nabla\cdot\left(H^\epsilon\,U^\epsilon\right)-\frac{1}{3}\Delta \zeta^\epsilon_t-\frac{1}{6}\;\Delta b^\epsilon_t=b^\epsilon_t,\vspace{0.3cm}\\
        U^\epsilon_t+\nabla \zeta^\epsilon +(U^\epsilon.\nabla)U^\epsilon=\mathbf{0},\vspace{0.3cm}\\
        \zeta^\epsilon(0,X)=0,\quad U^\epsilon(0,X)=\mathbf{0},
	\end{array}
	\right.
\end{equation}
with the bottom $b^\epsilon$  defined according to \eqref{bottom-type}. In the two classical tsunami approaches, this system is replaced by the following propagation system
\begin{equation}\label{Bous-Pd2}
	\left\{
	\begin{array}{ll}
		\zeta_t+\nabla\cdot\left(H\,U\right)-\frac{1}{3}\Delta \zeta_t=0,\vspace{0.3cm}\\
        U_t+\nabla \zeta +(U.\nabla)U=\mathbf{0},\vspace{0.3cm}\\
        \zeta(0,X)=\zeta_0(X),\quad U(0,X)=\mathbf{0}.
	\end{array}
	\right.
\end{equation}

 As mentioned in the introduction \ref{intro-paper}, the choice of $\zeta_0$ differs. Despite the fact that the passive approach is appropriate for the shallow-water system, this technique fails for \eqref{Bous-Pd2}. In this section, we rigorously establish that, when system \eqref{Bous-Pd1} is replaced by system \eqref{Bous-Pd2}, the appropriate initial free-surface displacement for tsunami propagation is given by (see \cite{chen2020mathematical} for the one-dimensional case):
 \begin{equation}\label{correct-intial-bousinsq}
     \widehat{\zeta_0}(k)=\frac{1-\frac{1}{6}|k|^2}{1+\frac{1}{3}|k|^2}\widehat{B}(k), \quad \forall k\in \mathbb{R}^2.
 \end{equation} 
 
Since the well-posedness theorem for systems \eqref{Bous-Pd1}-\eqref{Bous-Pd2} established in \cite{chen2020mathematical} is restricted to the one-dimensional setting and low-regularity Sobolev spaces, we first extend this result.

\begin{theorem}\label{uniqtheorem}
Let $s>2$ and $B\in H^{s+1}(\mathbb{R}^2)$ satisfy
\[
\inf_{\mathbb{R}^2}(1+\zeta_0-B)>0,
\]
where $\zeta_0$ is defined by \eqref{correct-intial-bousinsq}.
Assume that $f_\epsilon$, defined by \eqref{rising-type},
satisfies \eqref{hypotheses-type}. Then there exists
$T:=T(B,s)>0$, depending only on $B$ and $s$, such that
each of the systems \eqref{Bous-Pd1} and
\eqref{Bous-Pd2}
admits a unique solution $$\left(\underline{\zeta},\underline{U}\right) \in L^{\infty}\left(0,T; H^{s+1}(\mathbb{R}^2)\times H^s(\mathbb{R}^2;\mathbb{R}^2)\right)\bigcap C\left([0,T]; H^{s}(\mathbb{R}^2)\times H^{s-1}(\mathbb{R}^2;\mathbb{R}^2)\right).$$
Moreover, there exists a constant $R=R(B,s)>\left\Vert B\right\Vert_{H^{s+1}(\mathbb{R}^2)}$ depending only on $B$ and $s$, such that
$$\left\Vert (\underline{\zeta},\underline{U})\right\Vert_{L^{\infty}\left(0,T; H^{s+1}(\mathbb{R}^2)\times H^s(\mathbb{R}^2;\mathbb{R}^2)\right)}\leq R. $$
\end{theorem}

\begin{proof}
For ease of reading, the detailed proof is moved to Appendix \ref{sec:A}. Its core argument combines a fixed-point technique with some well-established results for transport equations \cite[Chapter 3]{bahouri2011fourier}.
\end{proof}
To rigorously analyze the error resulting from replacing \eqref{Bous-Pd1} by \eqref{Bous-Pd2}, where the initial tsunami condition is given by \eqref{correct-intial-bousinsq}, we introduce the following relative error:
\begin{equation}\label{Error-bous}
    E_s^{Bous}(t)=\frac{\left\Vert\zeta^\epsilon(t)-\zeta(t)\right\Vert_{H^s(\mathbb{R}^2)}+\left\Vert U^\epsilon(t)-U(t)\right\Vert_{H^{s-1}(\mathbb{R}^2;\mathbb{R}^2)}}{\left\Vert B\right\Vert_{H^{s}(\mathbb{R}^2)}},\quad s\geq 1,\quad\forall t\in (0,T),
\end{equation}
where $B\neq 0$. We are now in a position to state the second main result of the paper.
\begin{theorem}\label{convergence}
Let $s>2$ and assume that the hypotheses of Theorem \ref{uniqtheorem} are satisfied. Then there exists a constant $C:=C(T,R,s)$ , depending only on $T$, $R$ and $s$, such that the relative error defined by \eqref{Error-bous} satisfies
    \begin{equation}\label{upper-bound-error-Bous}
        E_s^{Bous}(t)\leq C\left\{  1-f_\epsilon(t)+ \left\Vert 1-f_\epsilon\right\Vert_{L^1(0,t) } \right\} ,\quad \forall t\in (0,T),
    \end{equation}
    where $f_\epsilon$ is given by \eqref{rising-type}. In particular, for all $t\in(t_\epsilon,T)$, we have

    \begin{equation}\label{after-ert-upper-bound-error-bous}
        E_{s_1}^{Bous}(t)\leq C_{Bous}\,t_\epsilon,\quad \forall s_1\in [1,s],
    \end{equation}
where $C_{Bous}$ depends on $C$ and $B$.
\end{theorem}

\begin{proof}
   Using the notation in \eqref{notation}, let $s>2$ and $B\in H^{s+1}$ be fixed. We choose  $T$ and $R$ as in Theorem~\ref{uniqtheorem}, and define $$V=U^{\epsilon}-U \quad \text{and} \quad  \xi=\zeta^{\epsilon}-\zeta.$$\\
    Subtracting the first equation in \eqref{Bous-Pd2} from that in \eqref{Bous-Pd1} and regrouping the terms, we obtain
    $$\xi(t)=-\displaystyle\int_0^t A\left(V+\zeta^{\varepsilon}V+\xi U-b^{\epsilon}V+(1-f_\epsilon)B\,U\right)(\tau)\,\dd \tau+ (f_{\epsilon}(t)-1)\zeta_0,$$
where $\zeta_0$ is given by \eqref{correct-intial-bousinsq} and $A$ by \eqref{oper-A}.  Then
\begin{equation}\label{Theorem3.2first-estimate}
    \left\Vert\xi(t)\right\Vert_{H^s}\leq 3(1+2 C_sR)\displaystyle\int_0^t\left[\left\Vert\xi\right\Vert_{H^s}+\left\Vert V\right\Vert_{H^{s-1}}\right](\tau)\,\dd \tau+ C_s R\left\Vert B\right\Vert_{H^s}\displaystyle\int_0^t (1-f_\epsilon(\tau))\;\dd \tau+ (1-f_{\epsilon}(t))\left\Vert B\right\Vert_{H^s}.
\end{equation}
Taking the difference of the second equations in systems \eqref{Bous-Pd1} and \eqref{Bous-Pd2} and regrouping the terms yields
\begin{equation}\label{format-second-eq}
    V_t+(U^{\epsilon}.\nabla)V+(\nabla U)V=-\nabla\xi
\end{equation}
Since $\left\Vert(\nabla U)V\right\Vert_{H^{s-1}}\leq C_s R\,\left\Vert V\right\Vert_{H^{s-1}}$, it follows from \cite[Theorem~3.14 and Remark 3.17]{bahouri2011fourier}  that there exists a constant $C(s) > 0$ such that
$$\left\Vert V(t) \right\Vert_{H^{s-1}}\leq \,e^{C(s)RT}\displaystyle\int_0^t\left\Vert\nabla\xi(\tau)\right\Vert_{H^{s-1}}\,\dd \tau\; .$$
Hence 
\begin{equation}\label{Theorem3.2-second-estimate}
    \left\Vert V(t) \right\Vert_{H^{s-1}}\leq \,e^{C(s)RT}\displaystyle\int_0^t\left\Vert\xi\right\Vert_{H^{s}}(\tau)\,\dd \tau\;.
\end{equation}
Summing \eqref{Theorem3.2first-estimate} and \eqref{Theorem3.2-second-estimate} term by term yields
\begin{equation}\label{inter-proof}
    \left\Vert V(t) \right\Vert_{H^{s-1}}+\left\Vert\xi(t)\right\Vert_{H^s}\leq C\displaystyle\int_0^t\left(\left\Vert V \right\Vert_{H^{s-1}}+\left\Vert\xi\right\Vert_{H^s}\right)(\tau)\;\dd\tau\;+C\left\Vert B\right\Vert_{H^s}(1-f_{\epsilon}(t))+C\left\Vert B\right\Vert_{H^s}\displaystyle\int_0^t(1-f_{\epsilon}(\tau))\,\dd \tau,
\end{equation}
where $C:= C(T,R,s) $  is a positive constant depending only on $T$, $R$ and $s$. \\
Setting 
\begin{equation}\label{inter-proof-2}
    \ell(t):=\displaystyle\int_0^t\left[\left\Vert V \right\Vert_{H^{s-1}}+\left\Vert\xi\right\Vert_{H^s}\right](\tau)\;\dd\tau, \quad \Theta(t):=\displaystyle\int_0^t (1-f_\epsilon(\tau))\,\dd \tau,
\end{equation}
 and multiplying inequality \eqref{inter-proof} by \(e^{-Ct}\), regrouping the terms, and using the notation introduced above \eqref{inter-proof-2}, we obtain the following equivalent inequality:
$$\frac{\dd}{\dd t}\left(\ell(t)e^{-Ct}\right)\leq C \left\Vert B\right\Vert_{H^s}e^{-Ct-t} \frac{\dd}{\dd t}\left( e^t \Theta(t)\right). $$
Integrating both sides, we obtain
$$\ell(t)\leq C \left\Vert B\right\Vert_{H^s} e^{CT+T} \Theta(t).$$
Substituting this inequality into \eqref{inter-proof} and regrouping the terms, we obtain
\begin{equation}\label{inter-proof-3}
    \left\Vert V(t) \right\Vert_{H^{s-1}}+\left\Vert\xi(t)\right\Vert_{H^s}\leq  \left\Vert B\right\Vert_{H^s} C\left\{ (1-f_{\epsilon}(t)) + \displaystyle\int_0^t(1-f_{\epsilon}(\tau))\,\dd \tau\right\} .
\end{equation}
Dividing both sides of the above inequality \eqref{inter-proof-3} by $\left\Vert B\right\Vert_{H^s}$ yields \eqref{upper-bound-error-Bous}.  Using arguments similar to those employed for \eqref{after-ert-upper-bound-error-NSW} in the proof of Theorem \ref{convergence-NSW}, we obtain the particular inequality \eqref{after-ert-upper-bound-error-bous}. This completes the proof.
\end{proof}
We now show that the solutions of \eqref{Bous-Pd1} coincide with those of \eqref{Bous-Pd2} in the case of sudden bottom motion. This result follows directly from Theorem \ref{uniqtheorem} and Theorem \ref{convergence}.
\begin{corollary}\label{Bous-equality}
Let $s>2$ and assume that the hypotheses of Theorem \ref{uniqtheorem} are satisfied. Suppose further that the bottom motion is instantaneous \eqref{insta_bottom}. Then the unique solution of
 \begin{equation}\label{Bous-Pd1-ins}
	\left\{
	\begin{array}{ll}
		\left(I-\frac{1}{3}\Delta\right)\zeta_t^0+\nabla\cdot\left(H^0\,U^0\right) = \delta_0\left(I+\frac{1}{6}\;\Delta\right) B,\vspace{0.3cm}\\
        U^0_t+\nabla \zeta^0 +(U^0.\nabla)U^0=\mathbf{0},\vspace{0.3cm}\\
        \zeta^0(0,X)=0,\quad U^0(0,X)=\mathbf{0},
	\end{array}
	\right.
\end{equation}
 is given by
\[
(\zeta^0(t),U^0(t))
=
\begin{cases}
(0,0), & t\leq 0,\\[1mm]
(\zeta(t),U(t)), & 0<t\leq T,
\end{cases}
\]
where $(\zeta,U)$ denotes the solution of  \eqref{Bous-Pd2}. In particular,
\[
\lim_{t\to 0^+}
\bigl(\zeta^0(t),U^0(t)\bigr)
=
(\zeta_0,0),
\]
where $\zeta_0$ is given by \eqref{correct-intial-bousinsq}. Moreover, for every $\delta\in(0,T)$, we have
\[
(\zeta^\epsilon,U^\epsilon)
\xrightarrow[t_\epsilon\to 0]{}
(\zeta^0,U^0)
\quad\text{in}\quad
C\!\left(
[\delta,T];
H^{s}(\mathbb{R}^2)
\times
H^{s-1}(\mathbb{R}^2;\mathbb{R}^2)
\right),
\]
where $(\zeta^\epsilon,U^\epsilon)$ denotes the
solution of \eqref{Bous-Pd1}.
\end{corollary}
To conclude this section, we give the corresponding results for the
linearized systems as in Section \ref{shalow}. The proof follows similar arguments to those in the proof of Theorem \ref{NSW_Estimate_linear}. 
\begin{theorem}\label{bous_Estimate_linear}
   Let $s\geq 0$ and $T>0$, and assume that $B\in H^{s+1}(\mathbb{R}^2)$. Then, for every $t\in(0,T)$ and $k\in \mathbb{R}^2$, the solutions of the linearized systems \eqref{Bous-Pd1} and \eqref{Bous-Pd2} are respectively given by
   \begin{equation}\label{pd1-S3-linear_solutions_bous}
	\left\{
	\begin{array}{ll}
		\widehat{\zeta^\epsilon}(t,k)=\left\{\displaystyle\int_0^t \cos\left(|\omega| (t-s)\right)f'_\epsilon(s)\;\dd s\right\}\widehat{\zeta_0}(k),\quad \widehat{U^\epsilon}(t,k)=-i\left\{\displaystyle\int_0^t \widehat{\zeta^\epsilon}(s,k) \,\dd s\right\} k,\vspace{0.3cm}\\
       \widehat{\zeta}(t,k)= \cos\left(|\omega|t\right)\widehat{\zeta_0}(k),\quad \widehat{U}(t,k)=-i\left\{\displaystyle\int_0^t \widehat{\zeta}(s,k) \,\dd s\right\} k, \\
       
	\end{array}
	\right.
\end{equation}
where $\widehat{\zeta_0}$ is given by \eqref{correct-intial-bousinsq} and $\omega^2(k)=\frac{|k|^2}{1+\frac{1}{3}|k|^2}$. Furthermore, for every $t\in(t_\epsilon,T)$, the  following estimates hold:
\begin{equation}\label{estimate_linear_bous}
    \left\Vert\zeta^\epsilon(t) -\zeta(t)\right\Vert_{H^{s}(\mathbb{R}^2)} \leq \sqrt{3}\, \left\Vert  B\right\Vert_{H^{s}(\mathbb{R}^2)}\,t_\epsilon
\end{equation}
and 
\begin{equation}\label{estimate_linear_velocity_bous}
    \left\Vert U^\epsilon(t) -U(t)\right\Vert_{H^{s-1}(\mathbb{R}^2)} \leq \left\Vert \nabla B\right\Vert_{H^{s-1}(\mathbb{R}^2)}\, t_\epsilon
\end{equation}
\end{theorem}

We conclude our theoretical study of the present work with the following
\begin{remark}\label{Bous_filter_Linear_filter} 
The Boussinesq and Kajiura Fourier multipliers agree through the quadratic term in the long-wave regime. Indeed,  we have  \[ \begin{aligned} \mathrm{KF}(k) &=\frac{1}{\cosh |k|}  =1-\frac{|k|^2}{2}+\frac{5|k|^4}{24}   +O(|k|^6),\\ \mathrm{BF}(k) &=\frac{1-|k|^2/6}{1+|k|^2/3}  =1-\frac{|k|^2}{2}+\frac{|k|^4}{6}   +O(|k|^6), \end{aligned} \] where $\mathrm{KF}$ and $\mathrm{BF}$ denote the Kajiura \eqref{intial_Euler} and Boussinesq \eqref{correct-intial-bousinsq} multipliers, respectively. This agreement is consistent with the long-wave regime $|k|\ll1$ in which the Boussinesq model is derived, see Section \ref{models_dis}. However, $\mathrm{BF}$ vanishes at $|k|=\sqrt6$, is negative for $|k|>\sqrt6$, and tends to $-1/2$ as $|k|\to\infty$. By comparison, $\mathrm{KF}$ is positive and decays exponentially. Thus, the term ``Boussinesq filter'' refers here to the initialization operator derived from the chosen Boussinesq system, whose agreement with the Euler initialization concerns
the long-wave limit. 
\end{remark}  
\begin{remark}
In the linear case, we can adopt a similar analysis to that in \cite{dutykh2007water,hammack1973,nosov1996,saito2013,saito2019tsunami} to derive closed-form expressions of the solutions for both the linear shallow-water and linear Boussinesq models and for different types of  bottom deformations (not only the separable case \eqref{bottom-type}); however, we omit this analysis here.
\end{remark}
\section{Numerical analysis}\label{numericalanal}
We divide this section into two subsections. In the first, we briefly present the  discretization that we employ to approximate both models \eqref{pd1} and \eqref{pd2}. In  Subsection \ref{sec:tests}, we perform numerical tests to validate our theoretical findings. For validation and comparison, we also plot the analytic solution of the linear potential equations, corresponding to the case of instantaneous bottom motion \eqref{insta_bottom} (see, e.g., \cite{hammack1973,kajiura1963leading}):
\begin{equation}\label{analytic_euler_solution}
       \zeta_{Euler}(t,X) = \frac{1}{(2\pi)^2} \int_{\mathbb{R}^2} \frac{\widehat{B}(k) \, e^{i\,k\cdot X}}{\cosh( |k|)} \cos(\omega_{Euler}t) \dd k,\quad \forall t>0,\,\forall X\in \mathbb{R}^2,
\end{equation}
where $\omega_{Euler}=\sqrt{|k|\tanh(|k|)}$.
\subsection{Discretization of the Models}\label{sec:dis}
For the time discretization, we adopt the second-order semi-implicit Crank--Nicolson--leap-frog  (CNLF) scheme used in \cite{chen2009numerical} to approximate the regularized Boussinesq system \cite{bona1998boussinesq}. For the spatial discretization, we employ a Galerkin finite element method with Q$_1$ Lagrange basis functions \cite{lucquin2004}. For further details on the numerical method and alternative choices of spatial discretization, we refer the reader to the main reference of this part \cite{chen2009numerical}.

We first introduce the following unified formulation for the shallow-water systems \eqref{pd1-S3}--\eqref{pd2-S3} and the Boussinesq systems \eqref{Bous-Pd1}--\eqref{Bous-Pd2}:

\begin{equation}\label{General_form}
	\left\{
	\begin{array}{ll}
		\zeta_t^\gamma+\nabla\cdot\left(H^\gamma\,U^\gamma\right)-\frac\gamma{3}\Delta \zeta^\gamma_t-\frac{\gamma}{6}\;\Delta b^\gamma_t=b^\gamma_t,\vspace{0.3cm}\\
        U^\gamma_t+\nabla \zeta^\gamma +(U^\gamma.\nabla)U^\gamma=\mathbf{0},\vspace{0.3cm}\\
        \zeta^\gamma(0,X)=\zeta_0^\gamma(X),\quad U^\gamma(0,X)=\mathbf{0}.
	\end{array}
	\right.
\end{equation}
Here, when $\gamma=0$, the formulation reduces to the shallow-water model, while $\gamma=1$ corresponds to the Boussinesq model. For a fixed value of $\gamma\in\{0,1\}$, the kinematic system is obtained by choosing
\[
b^\gamma=b^\epsilon:=f_\epsilon B,
\qquad
\zeta_0^\gamma=0.
\]
The corresponding static system is obtained by setting $f_\epsilon^{\gamma}(t)=1$ for all $t\geq0$. The initial condition $\zeta_0$ is then chosen according to the model: it is given by the bottom profile in the shallow-water case, and by the Boussinesq-filtered seabed \eqref{correct-intial-bousinsq} in the Boussinesq case.

For our algorithm, we consider the following equivalent formulation of \eqref{General_form}, obtained by introducing the change of variable $\eta^\gamma=\zeta^\gamma-b^\gamma$.
\begin{equation}\label{General_form_discr}
	\left\{
	\begin{array}{ll}
		(I-\frac\gamma{3}\Delta )\eta_t^\gamma+\nabla\cdot\left(H^\gamma\,U^\gamma\right)=\frac{\gamma}{2}\;\Delta b^\gamma_t,\vspace{0.3cm}\\
        U^\gamma_t +(U^\gamma.\nabla)U^\gamma=-\nabla \zeta^\gamma,\vspace{0.3cm}\\
        \eta^\gamma(0,X)=\zeta_0^\gamma(X)-b^\gamma(0,X),\quad U^\gamma(0,X)=\mathbf{0}.
	\end{array}
	\right.
\end{equation}
Let $\eta^{\gamma,n}$ and $U^{\gamma,n }$ denote the numerical approximations of $\eta^\gamma$ and $U^\gamma$ at time $t^n$. Following \cite{chen2009numerical}, we employ the CNLF scheme for the time discretization, which yields the following:

\begin{equation}\label{General_form_discr_time}
	\left\{
	\begin{array}{ll}
		(I-\frac{\gamma}{3}\Delta)P^n=-C^n \Delta t\left[ \nabla\cdot(H^\gamma U^\gamma)\right]^n +\frac{\gamma}{2}f_\epsilon^{\gamma,n} \Delta B,\vspace{0.3cm}\\
        Q^n=-C^n \Delta t\left[ \left(U^\gamma\cdot \nabla\right)U^\gamma+\nabla\zeta^\gamma \right]^n, \vspace{0.3cm}\\
        \zeta^{\gamma,n+1}= \eta^{\gamma,n+1}+f_\epsilon^{\gamma,n+1} B,
	\end{array}
	\right.
\end{equation}
for $n=0,1,2, \dots$,  where 
$$P^n=\left\{
	\begin{array}{ll}
      \eta^{\gamma,1}-\eta^{\gamma,0}, \, n=0,\\
      \eta^{\gamma,n+1}-\eta^{\gamma,n-1}, \, n\geq 1,\\
    \end{array}
    \right. , \quad Q^n=\left\{
	\begin{array}{ll}
      U^{\gamma,1}-U^{\gamma,0}, \, n=0,\\
      U^{\gamma,n+1}-U^{\gamma,n-1}, \, n\geq 1,\\
    \end{array}
    \right. , \quad f_\epsilon^{\gamma,n}=\left\{
	\begin{array}{ll}
      f_\epsilon^{\gamma,1}-f_\epsilon^{\gamma,0}, \, n=0,\\
      f_\epsilon^{\gamma,n+1}-f_\epsilon^{\gamma,n-1}, \, n\geq 1,\\
    \end{array}
    \right.
    $$
and 
$$C^n=\left\{
	\begin{array}{ll}
      1, \, n=0,\\
      2, \, n\geq 1,\\
    \end{array}
    \right.  ,\quad n\geq 0.
    $$

At a fixed time $t^n$, let $R_1^n$ and $\bm{R}_2^n$ denote the right-hand sides of the first and second equations in \eqref{General_form_discr_time}, respectively. The computation of $\eta^{\gamma,n+1}$ and $U^{\gamma,n+1}$ then reduces to solving the following problem:
\begin{equation}\label{Elliptic_sys}
    \left\{
	\begin{array}{ll}
    (I-\frac{\gamma}{3}\Delta)P^n=R^n_1,\vspace{0.3cm}\\
     Q^n=\bm{R}_2^n.
    \end{array}
    \right.
\end{equation}
Let
$ 
\Omega=(0,L_x)\times(0,L_y)
$  be the nondimensional domain of interest, where $L_x$ and $L_y$ denote its length and width, respectively. For the study of tsunami propagation in $\Omega$, we impose the following homogeneous Dirichlet boundary conditions:
\begin{equation}\label{Elliptic_boundary}
    \left\{
	\begin{array}{ll}
    \eta^\gamma=0,\quad \partial \Omega,\vspace{0.1cm}\\
    U^\gamma =\bm{0},\quad \partial \Omega.
    \end{array}
    \right.
\end{equation}
As mentioned at the beginning of this subsection, we employ a Q$_1$ Lagrange finite-element method to discretize system \eqref{Elliptic_sys}, subject to the homogeneous Dirichlet boundary conditions \eqref{Elliptic_boundary}, with a prescribed spatial discretization step $\Delta x=\Delta y=\frac{L_x}{N_x}$. The right-hand-side integrals, including the linear free-surface gradient and the nonlinear terms, are evaluated using a $2\times2$  Gauss–Legendre quadrature rule on each  element, and the initial free-surface condition for the static Boussinesq system is approximated as
\begin{equation*}
    \left(M +\frac{1}{3} K\right)\eta^0_h = -\frac{1}{2} K\,B_h,
\end{equation*}
where $M$ and $K$ are the mass and stiffness matrices, respectively.
Moreover, we solve the linear systems using a fast Laplace solver (DST-I),
with computational complexity $\mathcal{O}(N_x^2\log(N_x^2))$
\cite{stefano2000}.

For completeness, we conclude this part with the following.

\begin{remark}
Higher-order accuracy for the elliptic system \eqref{Elliptic_sys}-\eqref{Elliptic_boundary} can be achieved using spectral methods \cite{shen1994efficient} or isogeometric discretizations \cite{lamsahel2025outlier}. However, the present discretization is sufficient for validating Theorem \ref{convergence-NSW} and Theorem \ref{convergence}.
\end{remark}

\subsection{Numerical tests}\label{sec:tests}
We consider a propagation domain $\Omega$ with length and width both equal to $450$ km. For brevity, all tests are performed using the vertical Okada dip-slip fault bottom. Similar results are obtained for other bottom profiles, including those in \cite{chen2009numerical,dutykh2007water,poletto2025full}. 

\begin{figure}[htbp]
    \centering
    \small
    \begin{minipage}[c]{0.42\textwidth}
        \centering
        \footnotesize
        \renewcommand{\arraystretch}{0.85}
        \begin{tabular}{lc}
        \hline
        \textbf{Parameter} & \textbf{Value} \\
        \hline
        Fault depth (km) & 10 \\
        Fault length (km) & 20 \\
        Fault width (km) & 16\\
        Burger's vector length (m) & 6 \\
        Water depth, $H_0$ (km) & 3 \\
        Young modulus (GPa) & 9.5 \\
        Poisson ratio & 0.27 \\
        Dip angle ($^\circ$) & 13 \\
        Strike angle ($^\circ$) & 90 \\
        Normal angle ($^\circ$) & 0 \\
        gravitational acceleration, $g$ (m/s$^2$) & 9.812 \\
        \hline
        \end{tabular}
        \captionof{table}{{\footnotesize Fault parameters considered in the numerical tests (see \cite{dutykh2007water} for details).}}
        \label{tab:parameters}
    \end{minipage}
    \hfill
    \begin{minipage}[c]{0.52\textwidth}
        \centering
        \includegraphics[width=\textwidth]{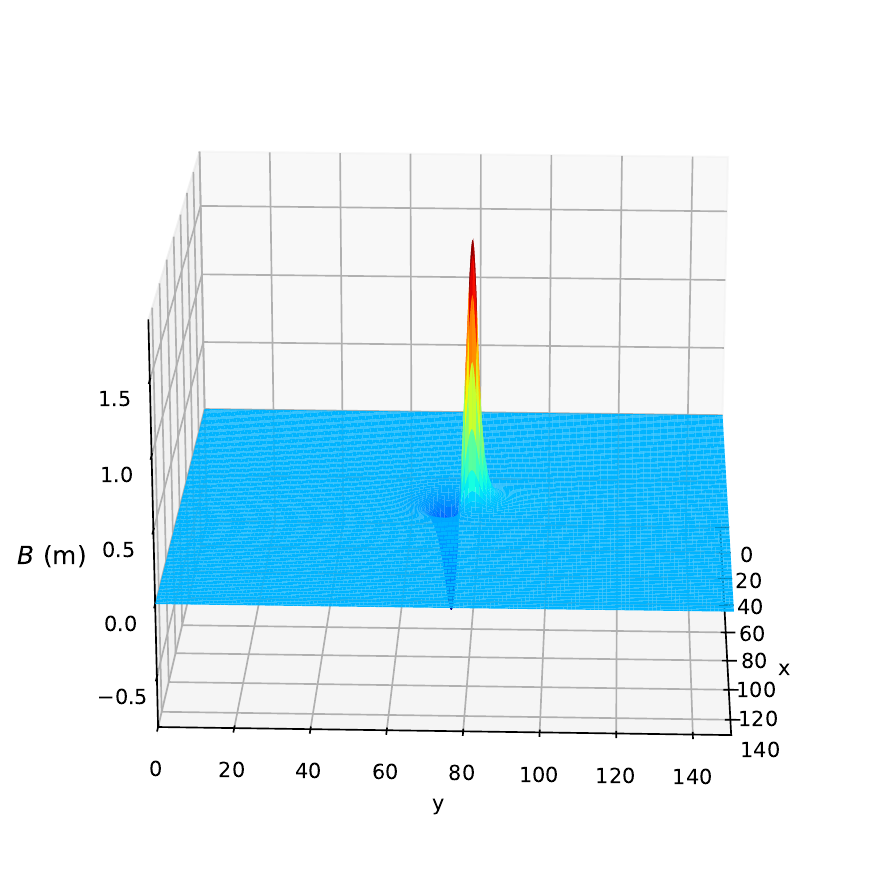}
        \captionof{figure}{ {\footnotesize Vertical seabed deformation generated by an idealized rectangular dip-slip fault with the parameters listed in Table ~\ref{tab:parameters}. The horizontal coordinates are dimensionless.}}
        \label{fig: 2}
    \end{minipage}
\end{figure}

We assume that the static bottom deformation $B$ in \eqref{bottom-type}  can be described by the analytical Okada solutions  \cite{okada1985surface,okada1992internal}; see also \cite[Section 2]{dutykh2007water} and the references therein. In particular, we consider the so-called vertical dip-slip dislocation fault; see \cite{mitsotakis2009boussinesq}. The corresponding fault parameters used in our numerical tests are listed in Table~\ref{tab:parameters}, while Figure \ref{fig: 2} illustrates the resulting bottom profile,  with the fault placed at the center of the computational domain.

For static systems \eqref{pd2-S3} and \eqref{Bous-Pd2}, the  bottom profile shown in Figure~\ref{fig: 2} is sufficient for the numerical analysis of tsunami propagation. In contrast, kinematic systems \eqref{pd1-S3} and \eqref{Bous-Pd1}  require a rise function \eqref{rising-type}. The choice of this function can affect the amplitude of the waves generated; see \cite{dutykh2007water,sementsov2025effect}. In the following numerical tests, we therefore consider the trigonometric earthquake-rise function \cite{hammack1973,poletto2025full}:
\begin{equation}\label{choosed_type}
    f_{t_\epsilon}(t)=\frac{1}{2}\left(1-\cos\left(\frac{\pi t}{t_\epsilon}\right) \right),\quad \forall t\in (0,t_\epsilon).
\end{equation}

\begin{figure}[!htb]
  \centering
  \includegraphics[width=0.9\textwidth]{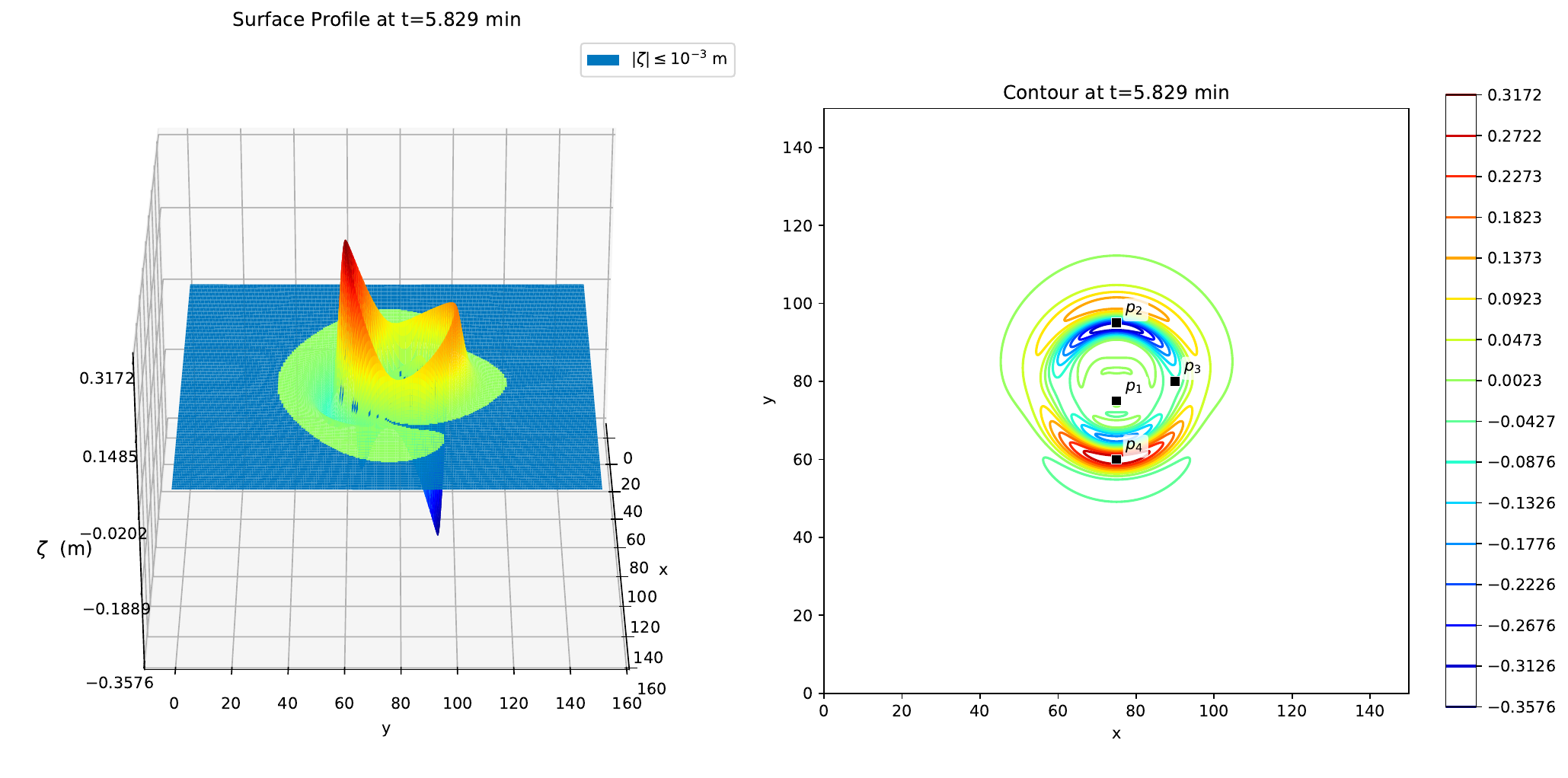}
  \caption{{\footnotesize Tsunami waves generated by the static Boussinesq system \eqref{Bous-Pd2}--\eqref{correct-intial-bousinsq}.}}
  \label{fig:surafce1}
\end{figure}

It is important to recall that we are approximating the nondimensional systems. Nevertheless, the relation between the dimensional and nondimensional parameters is given in Remark~\ref{nondimensional_change}. Throughout the tests, we use the following  parameters in the numerical algorithm \eqref{General_form_discr_time}--\eqref{Elliptic_boundary}:
\begin{equation}\label{Dx_Dt}
     \Delta t=\frac{20}{3000}\approx 0.0067,\quad \Delta x=\frac{150}{800}=0.1875\;.
\end{equation}
In order to measure the error at specific artificial tide-gauge stations, we introduce the following local relative error \cite{dutykh2006linear}.
\begin{equation}\label{local_error}
    \mathcal{R}(t,X)=\frac{\left|\zeta^{\epsilon}(t,X)-\zeta(t,X)\right|}{\max_{t_\epsilon\leq \tau \leq T} \left|\zeta(\tau,X)\right|},\quad \forall t\in (t_\epsilon,T),\;\;\forall X \in \mathbb{R}^2.
\end{equation}
Finally, it should be mentioned that the final time $T=20$
is chosen such that the waves remain sufficiently far from
the boundary; see Figure \ref{fig:surafce1}. Moreover, the analytical solutions
\eqref{pd1-S3-linear_solutions}, \eqref{pd1-S3-linear_solutions_bous}, and
\eqref{analytic_euler_solution}, together with the relative errors \eqref{relativeerror} and \eqref{Error-bous}, are evaluated numerically using the fast Fourier transform (FFT).\\

\noindent  {\bf \em Test 1: The impact of the passive  approach on the static Boussinesq system } \vspace*{0.25cm}
$\quad$\\
Before proceeding to the numerical validation of our theoretical results, we first illustrate the impact of adopting the passive-generation approach for the static Boussinesq system \eqref{Bous-Pd2}. Unless otherwise stated, time and the horizontal coordinates in the subsequent plots are nondimensional, whereas the surface elevation is given in meters (see Remark \ref{nondimensional_change}).

Figure \ref{fig:surafce1} shows the tsunami waves after $5.829$ min, obtained using the static system \eqref{Bous-Pd2} with the theoretical initial condition \eqref{correct-intial-bousinsq}. The waves exhibit significantly larger amplitudes in the $y$-direction than in the $x$-direction. This directional difference results from the chosen fault orientation, as illustrated in Figure~\ref{fig: 2}. Further discussion of this phenomenon can be found in \cite{chen2009numerical,grilli2007source}.

\begin{figure}[!htb]
    \centering
    \includegraphics[width=0.43\textwidth]{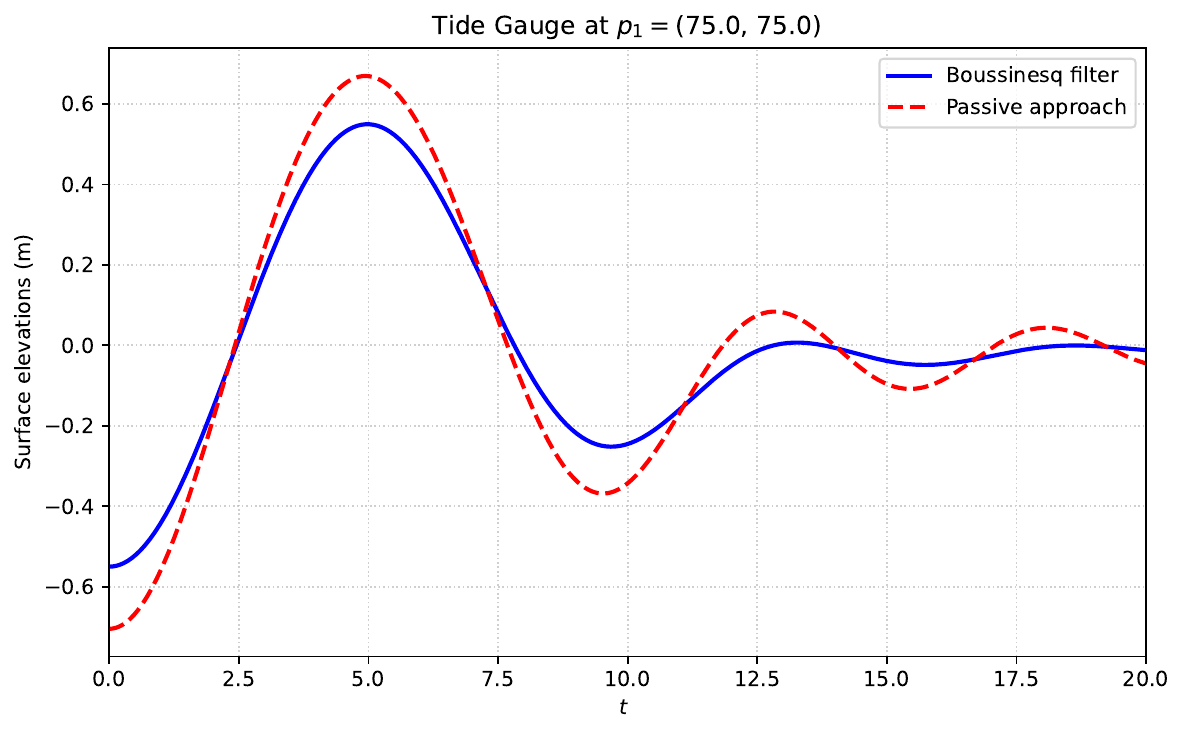}\hspace{1em}%
    \includegraphics[width=0.43\textwidth]{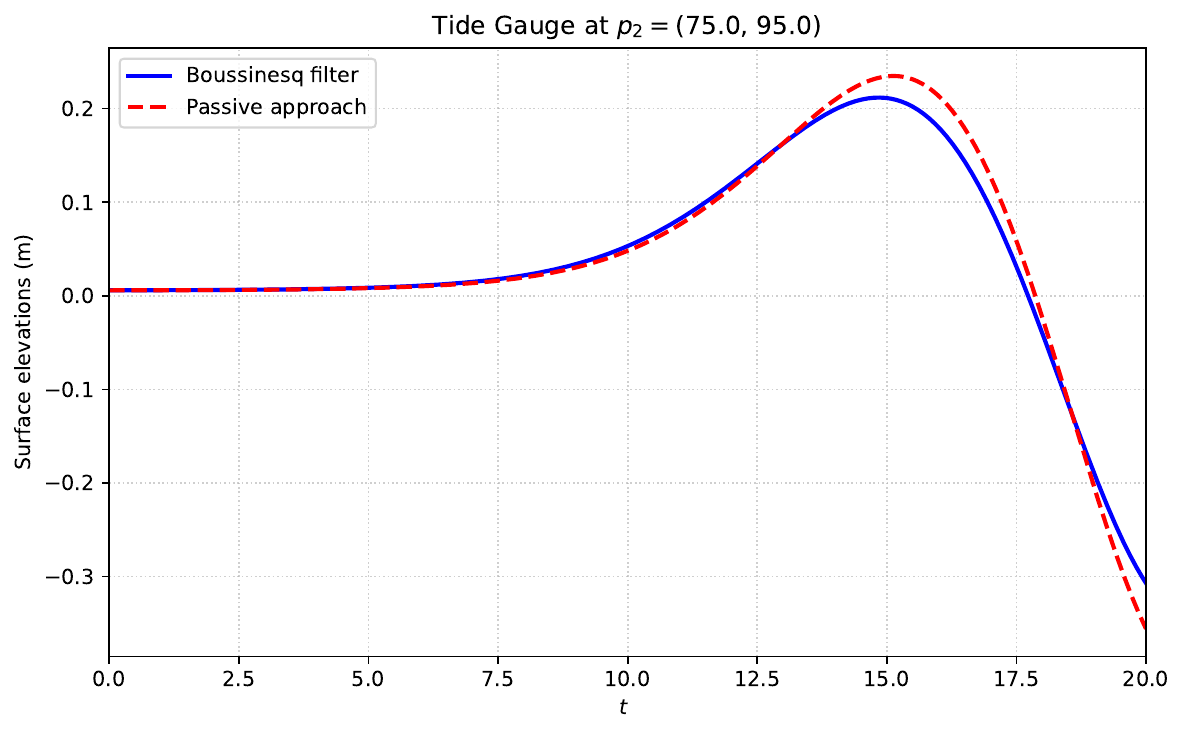}
    
    \vspace{0.5em} 
    
    \includegraphics[width=0.43\textwidth]{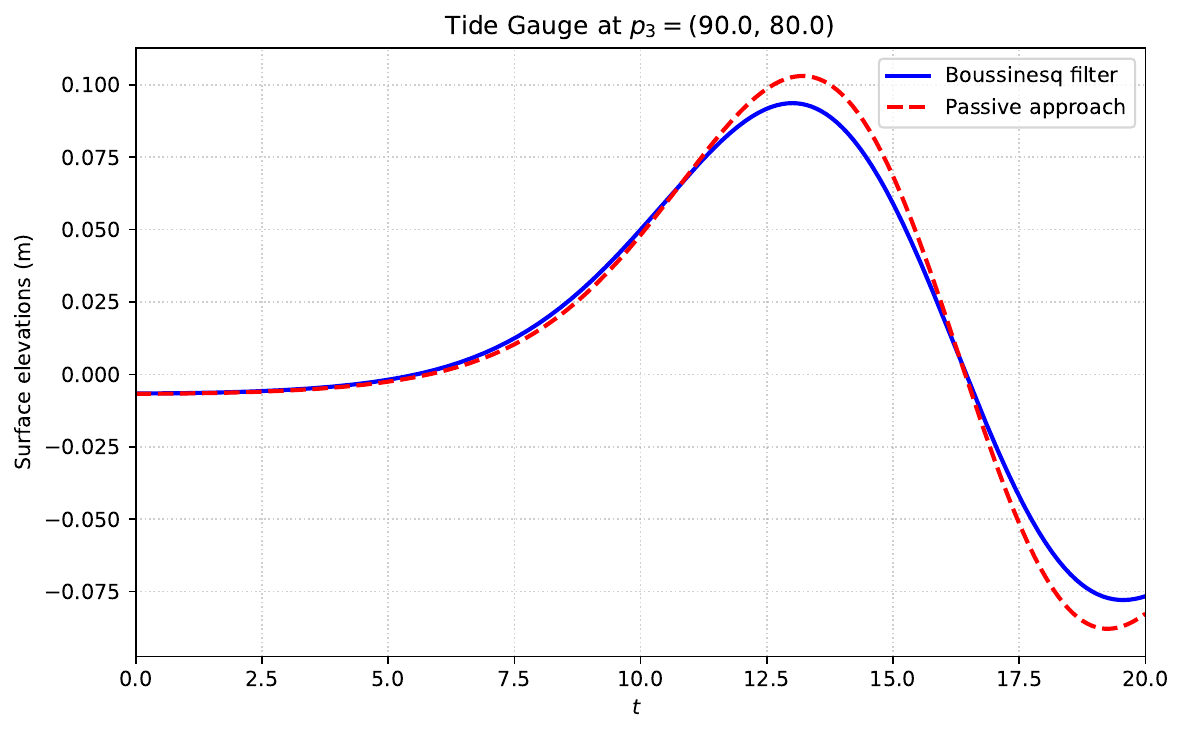}\hspace{1em}%
    \includegraphics[width=0.43\textwidth]{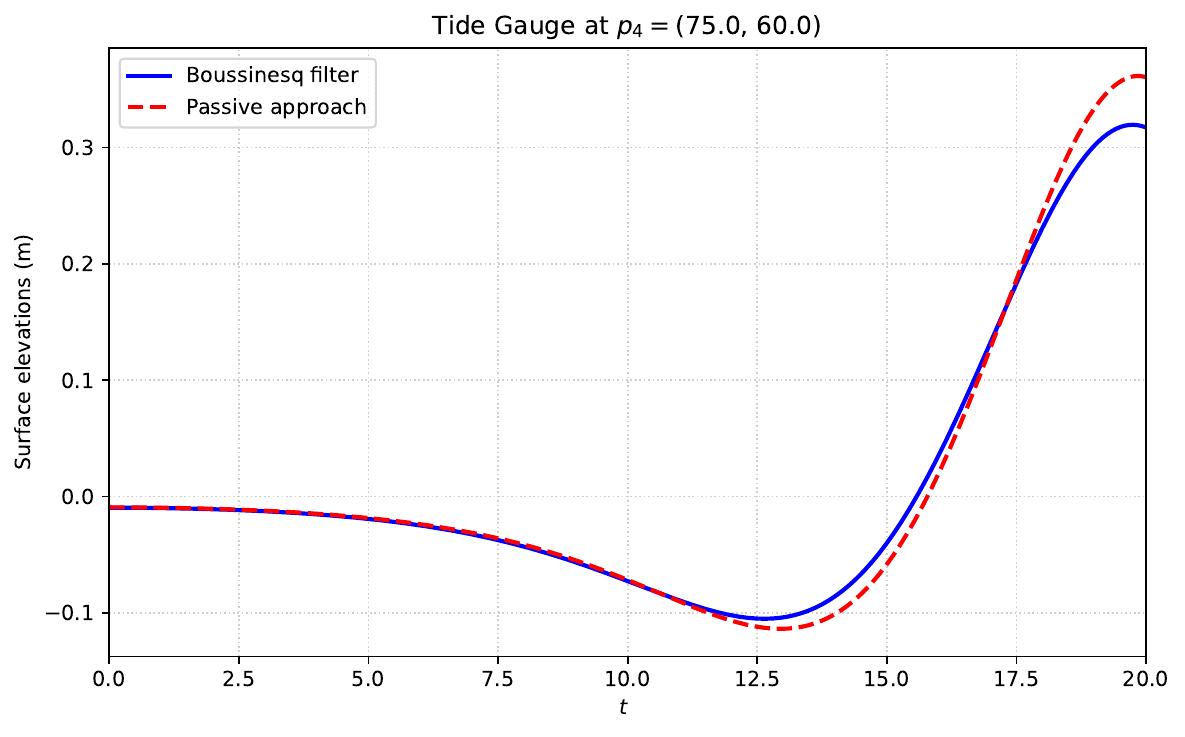}
    
    \caption{{\footnotesize Time series of the two surface elevations at selected tide-gauge locations.}}
    \label{fig:tide_gauges_all}
\end{figure}

We now follow \cite{dutykh2006linear} to investigate the error introduced by the passive approach. Figure~ \ref{fig:tide_gauges_all} compares the surface profiles obtained using passive generation with those obtained using the Boussinesq-filtered bottom \eqref{correct-intial-bousinsq} at several locations (see Figure  \ref{fig:surafce1}). The results show that  the waves generated by the passive approach exhibit noticeably larger amplitudes than those obtained using \eqref{correct-intial-bousinsq}. More precisely, we observe that  $\zeta_{\mathrm{passive}}$ is larger than $\zeta$ over the concave regions of $\zeta$ and smaller over the convex regions. Similar behavior is found  for the linear potential model; see \cite{dutykh2006linear}.

The local relative error between the two profiles ($\left|\zeta_{\mathrm{passive}}(t,X)-\zeta(t,X)\right|/\max_{0\leq \tau \leq T} \left|\zeta(\tau,X)\right|$) at the selected locations ranges from 13\% at $p_3$ to 28\% at $p_1$. Although this error depends on the specific problem configuration, including the water depth and bottom
topography, it remains significant for all test cases considered.

\begin{figure}[!htb]
    \centering
    \includegraphics[width=0.43\textwidth]{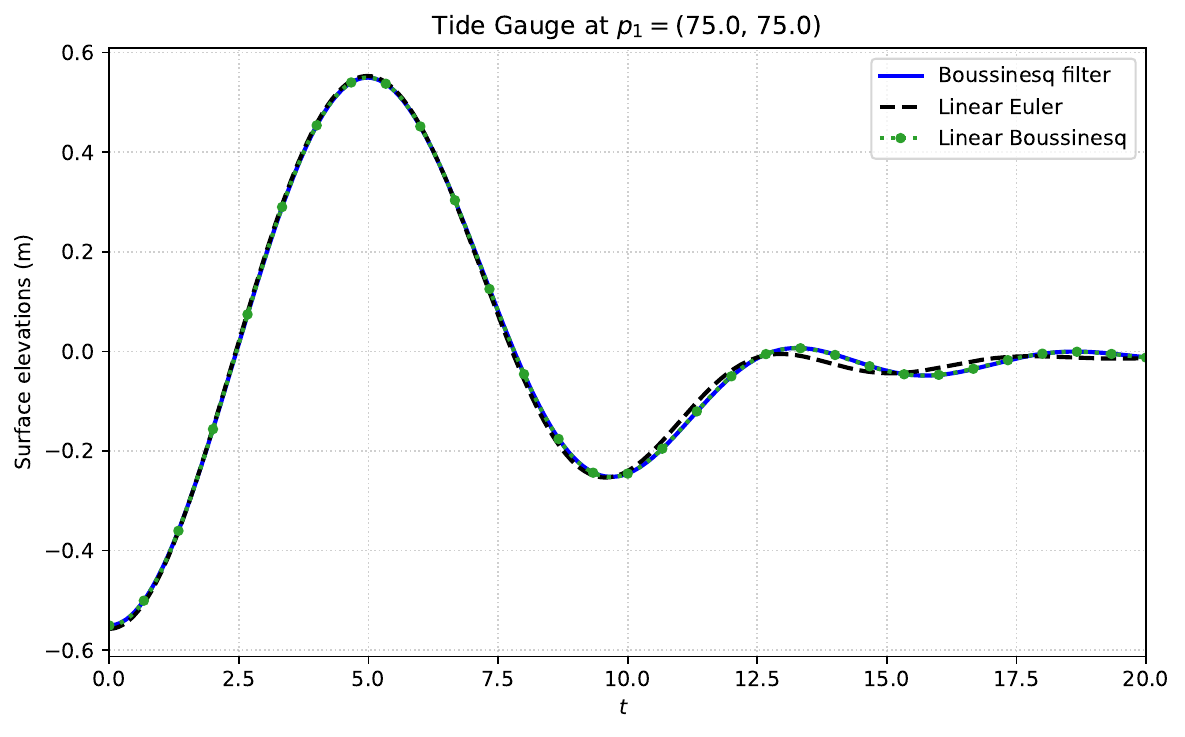}\hspace{1em}%
    \includegraphics[width=0.43\textwidth]{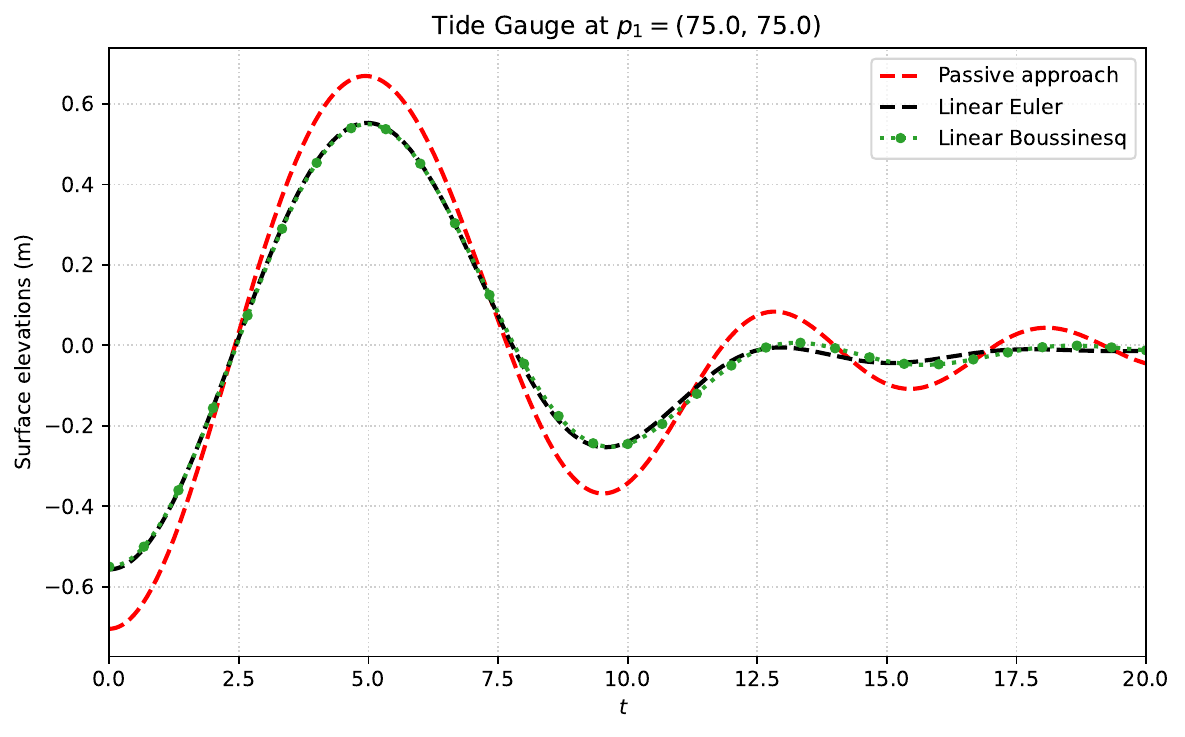}
    
    \vspace{0.5em} 
    
    \includegraphics[width=0.43\textwidth]{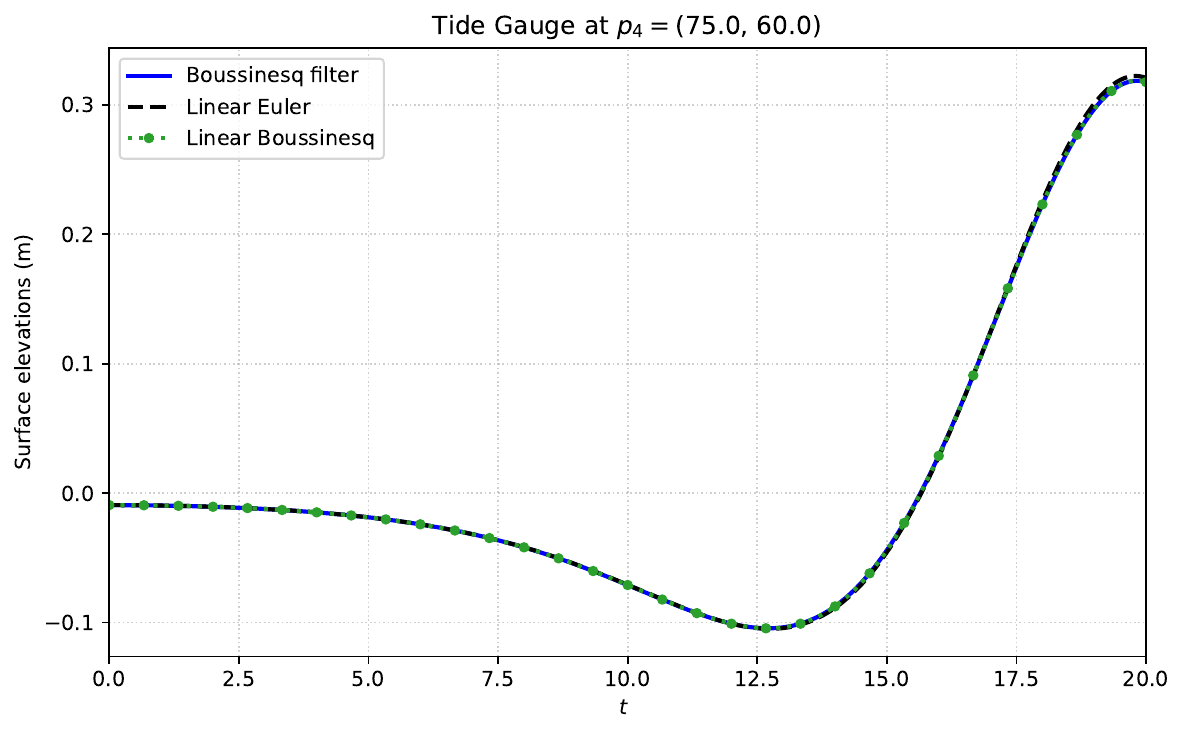}\hspace{1em}%
    \includegraphics[width=0.43\textwidth]{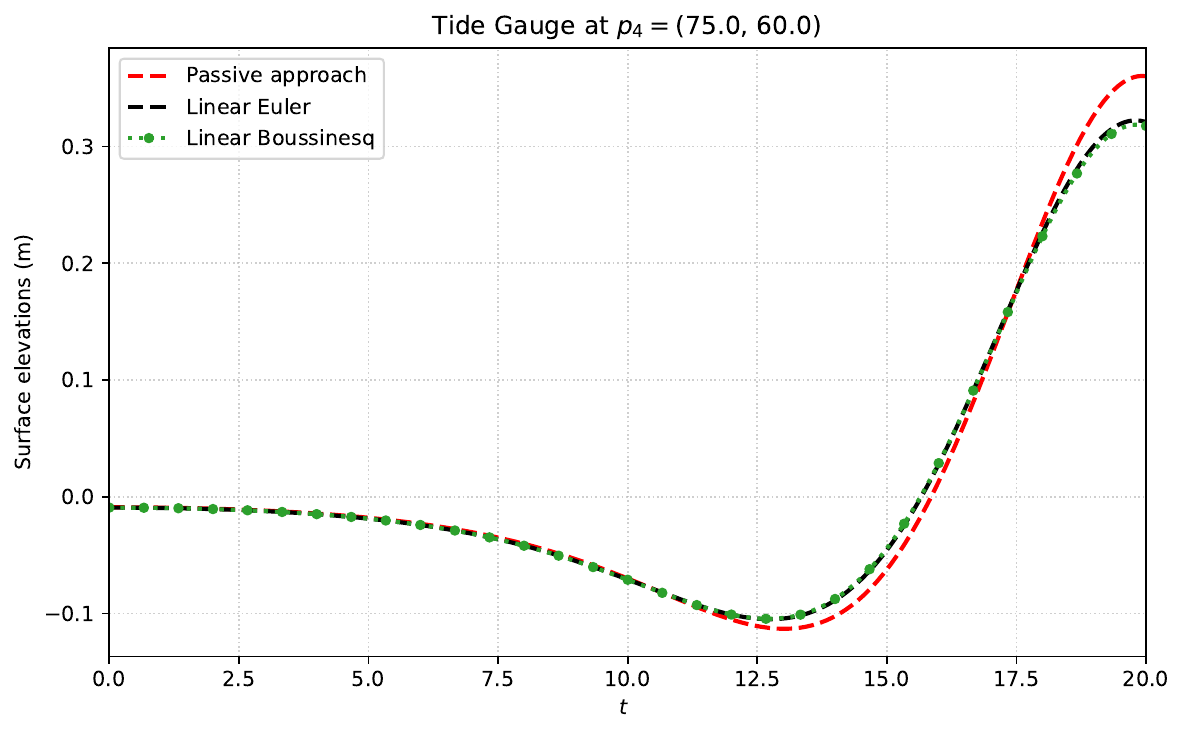}
    
    \caption{{\footnotesize Boussinesq-filtered static and passive static solutions vs. linear Euler and linear Boussinesq solutions at two nondimensional tide-gauge locations.}}
    \label{fig:Eulercompa}
\end{figure}

To better illustrate the error introduced by the passive generation  and to further validate our model with  initial condition \eqref{correct-intial-bousinsq}, Figure~\ref{fig:Eulercompa} compares the generated surface elevations at two locations with the corresponding linear Euler solution \eqref{analytic_euler_solution} and the linear Boussinesq solution \eqref{pd1-S3-linear_solutions_bous}. The results show  good agreement between our model and the linear potential solution \eqref{analytic_euler_solution}, while the passive approach clearly  overestimates the elevation of the surface. Furthermore, the numerical solution agrees  with the linear Boussinesq solution \eqref{pd1-S3-linear_solutions_bous}, as expected for a very small nonlinearity parameter
$\varepsilon\approx 10^{-4}$. This, in particular, provides a validation
of our numerical results. However, it should be emphasized that this agreement between the linear Boussinesq solution and our solver is limited to the short propagation time $T$ and the large water depth $H_0$ (weakly nonlinear regime), see, e.g.,  \cite{dutykh2013finite} for details. \\


\noindent  {\bf \em Test 2: Validation of the theoretical results of Section~\ref{bous} } \vspace*{0.25cm}
$\quad$\\
Having highlighted the importance of the appropriate initial condition \eqref{correct-intial-bousinsq}, we are now in a position to validate the convergence of the kinematic system \eqref{Bous-Pd1} toward the static system \eqref{Bous-Pd2} as the earthquake rise time decreases. 

\begin{figure}[!htb]
    \centering
    \includegraphics[width=0.43\textwidth]{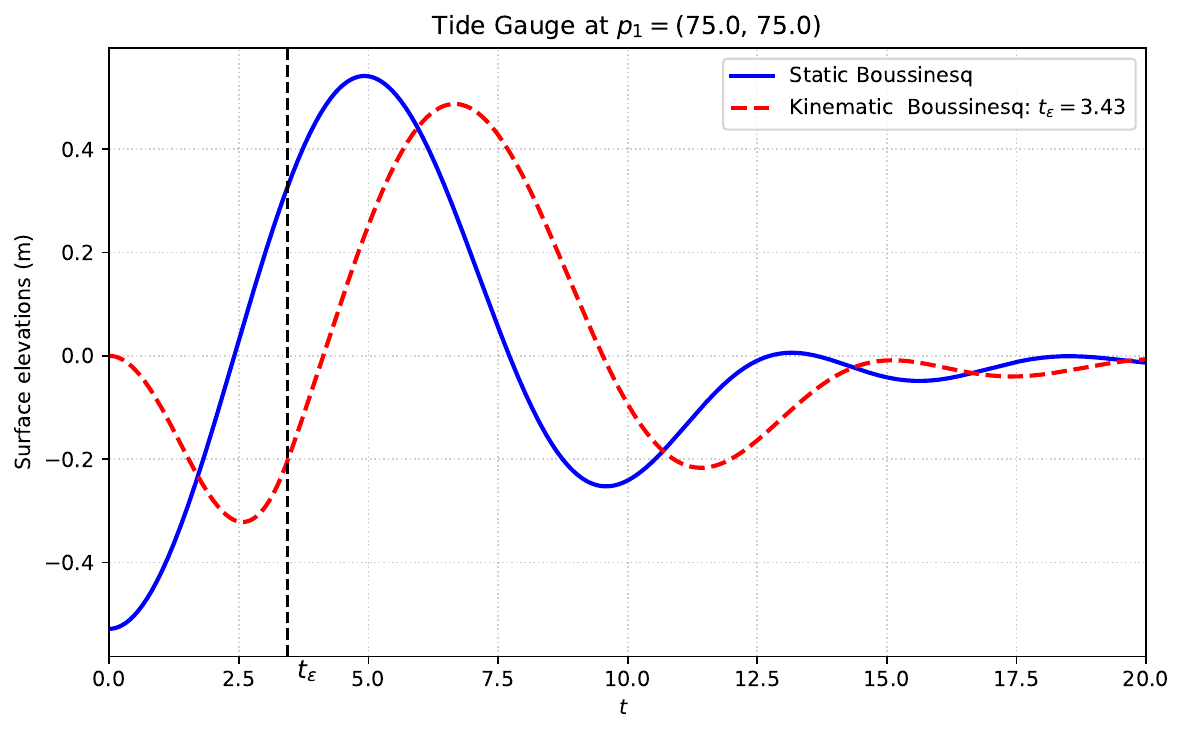}\hspace{1em}%
    \includegraphics[width=0.43\textwidth]{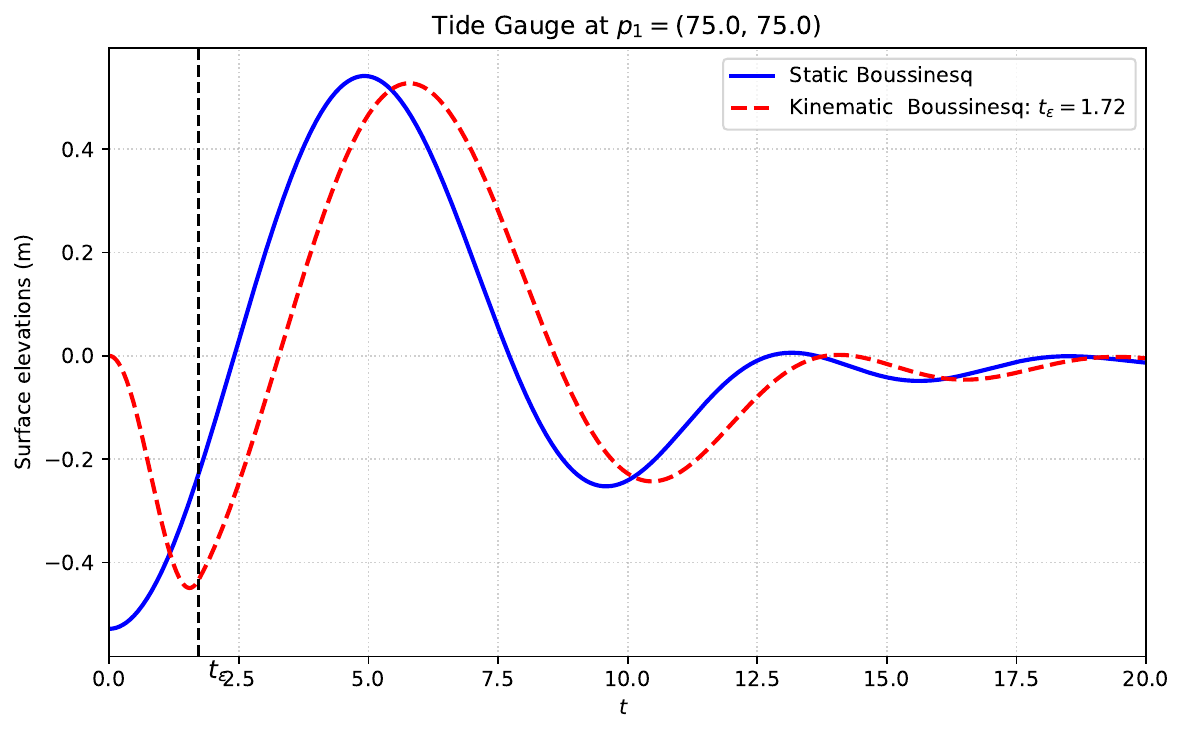}
    
    \vspace{0.5em} 
    
    \includegraphics[width=0.43\textwidth]{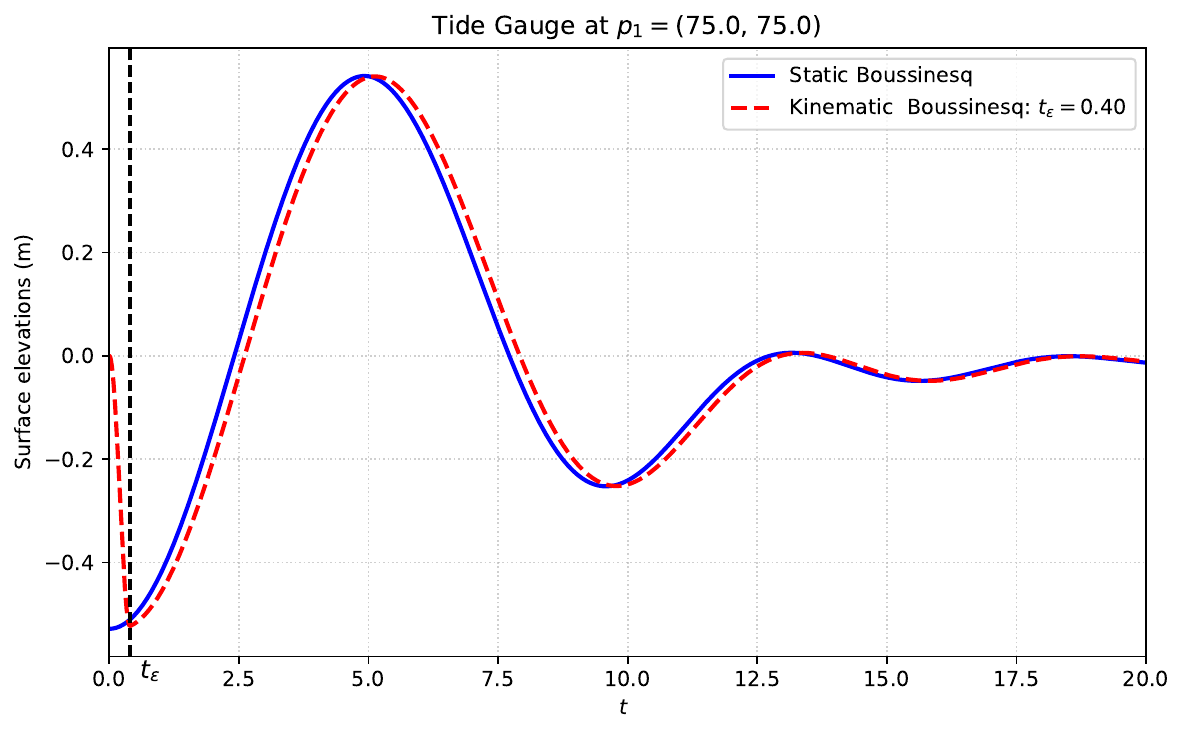}\hspace{1em}%
    \includegraphics[width=0.43\textwidth]{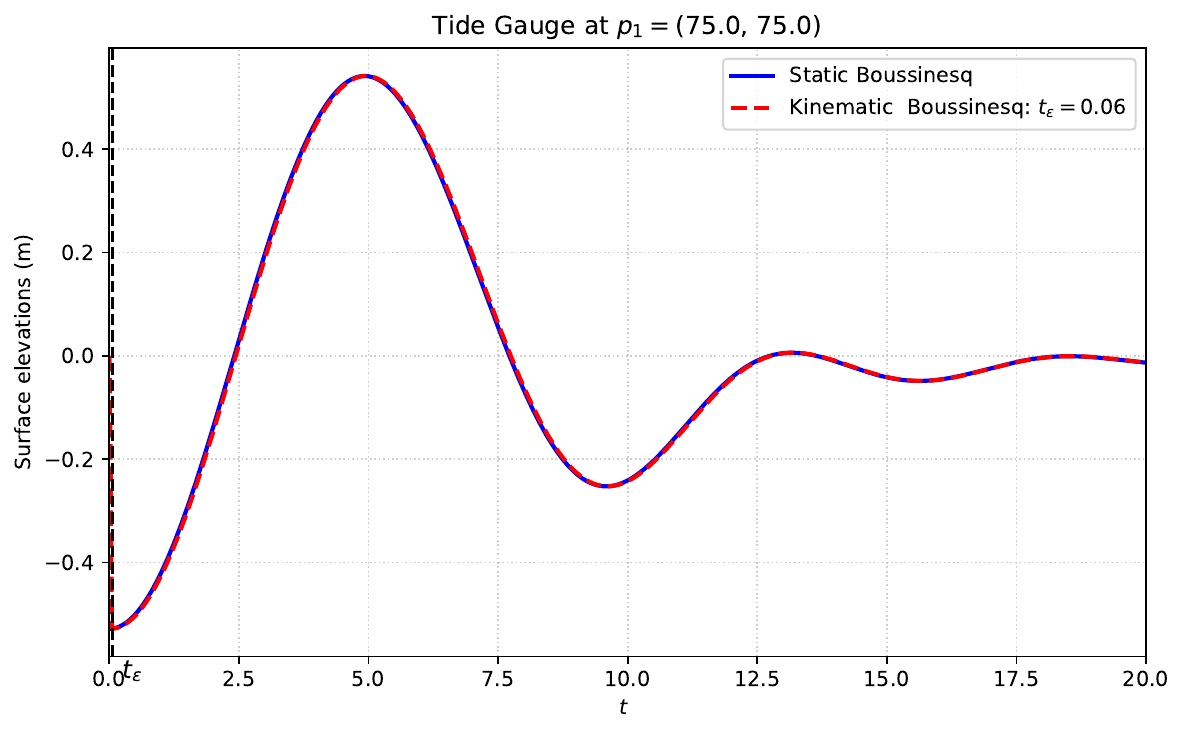}
    
    \caption{{\footnotesize Surface elevations for kinematic and static systems at $p_1=(75,75)$ with different rise times $t_\epsilon$.}}
    \label{fig:convergenceP1}
\end{figure}

Figure~\ref{fig:convergenceP1} shows the surface elevations of the kinematic and static systems at position $p_1=(75,75)$ for different rise times $t_\epsilon$, corresponding to physical rise times of 1 min, 30 s, 7 s, and 1 s, respectively. Except in the bottom-motion interval $(0, t_\epsilon)$, we clearly observe the convergence of the kinematic wave towards the static wave as $t_\epsilon$ decreases, which is consistent with the convergence predicted by Theorem \ref{convergence}.  Moreover, for relatively large values of $t_\epsilon$, the kinematic system produces waves of smaller amplitude and with noticeably delayed arrival times compared with
the static system; see also \cite[Figure 1]{sementsov2025effect} for a similar
observation in the linear Euler framework. Note that similar results are  observed for other locations as well.

\begin{figure}[!htb]
    \centering
    \includegraphics[width=0.45\textwidth]{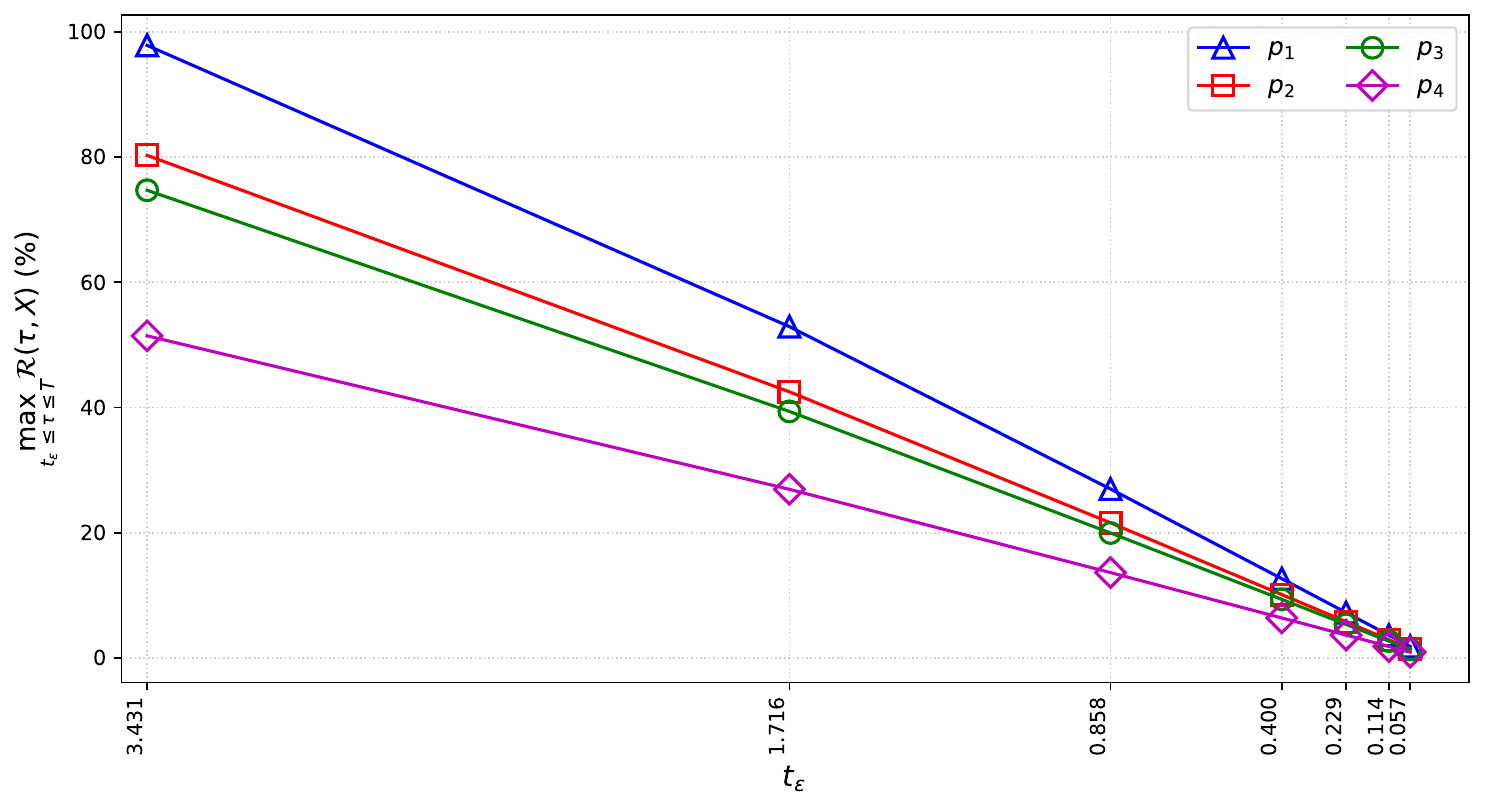}\hspace{1em}%
    \includegraphics[width=0.45\textwidth]{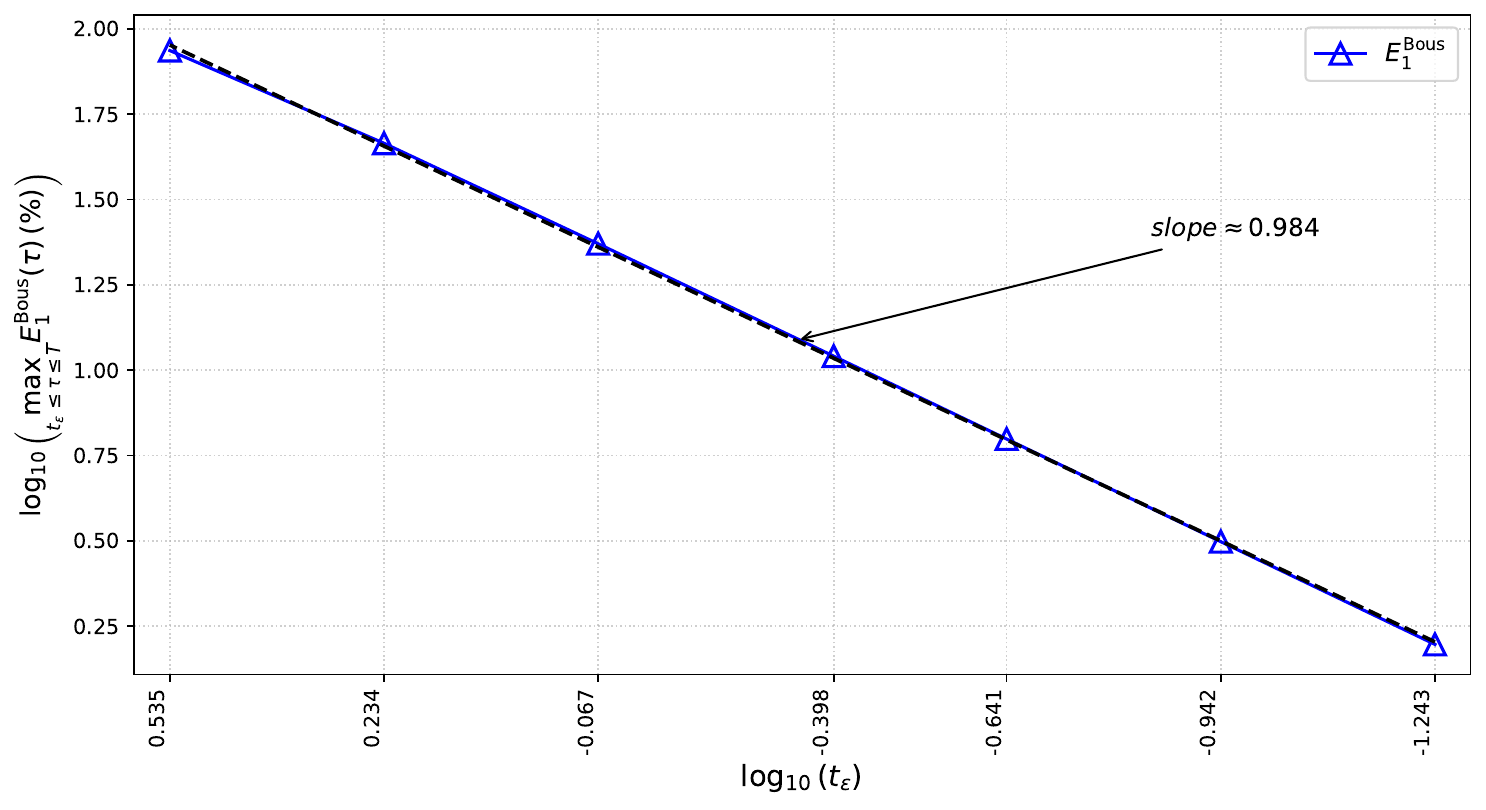}

    \caption{{\footnotesize Left: Maximum pointwise relative error $\mathcal{R}$ at selected locations $p_i$ versus the earthquake rise time $t_\epsilon$. Right: $\log_{10}$ of  the maximum Boussinesq error $E_1^{Bous}$ versus $\log_{10}(t_\epsilon)$. In both panels the  maximum is taken over $\tau \in (t_\epsilon, 20)$.}}
    \label{fig:relativeerrorbous}
\end{figure}

To quantify the computed error between the two surfaces, we show in the left panel of Figure~\ref{fig:relativeerrorbous} the local relative error values \eqref{local_error} for the chosen locations (see Figure~\ref{fig:surafce1}) over a range of earthquake rise times corresponding to $\{60,30,15,7,4,2,1 \}\,\text{s}$. The local relative error stays above 50\% when the rise time is 1 min, with a maximum of 97.8\% recorded at location $p_1$. Nevertheless, the error decreases nearly linearly with decreasing rise time $t_\epsilon$. For example,  at location $p_1$, the error drops from 97.8\% at 1 min to 3.6\% at 2~s, and further to 1.8\% at 1~s. In the right panel of Figure~\ref{fig:relativeerrorbous}, we plot the relative error $E_s^{\mathrm{Bous}}$ given by \eqref{Error-bous} for $s=1$. Similar to the local error, the error $E_s^{\mathrm{Bous}}$ is observed to decrease almost linearly, with an upper bound constant of approximately  0.3. This agrees with the estimate \eqref{after-ert-upper-bound-error-bous}.

To measure the sensitivity of the computed error \(E_1^{\mathrm{Bous}}\) to temporal and spatial refinement, Table \ref{tab:dx_dt_E_bous} reports its values for a rise time of \(1\,\mathrm{s}\). The results show that the computed error remains essentially unchanged under refinement. In particular, with \(N_x=800\) fixed, increasing \(N_t\) from 3000 to 6000 changes the error by approximately \(0.013\%\). With \(N_t=6000\) fixed, increasing \(N_x\) from 400 to 800 changes it by approximately \(0.11\%\).

\begin{table}[ht]
    \centering
    \renewcommand{\arraystretch}{1.2}
    \begin{tabular}{|c|c|c|c|}
        \hline
        \diagbox{$N_x$}{$N_t$}
        & $1000$ & $3000$ & $6000$ \\ \hline
        $200$ &  $1.5702\times 10^{-2}$ & $1.5645\times 10^{-2}$ & $1.5643\times 10^{-2}$ \\ \hline
        $400$ & $1.5770\times 10^{-2}$ & $1.5713\times 10^{-2}$ & $ 1.5711\times 10^{-2}$ \\ \hline
        $800$ & $1.5786\times 10^{-2}$ & $1.5730\times 10^{-2}$ & $1.5728\times 10^{-2}$ \\ \hline
    \end{tabular}
    \caption{{\footnotesize Sensitivity of the error $E_1^{Bous}$ shown in
    Figure \ref{fig:relativeerrorbous} to temporal ($\Delta t=20/N_t$)
    and spatial refinement, for a rise time $t_\epsilon$ of 1 s.}}
    \label{tab:dx_dt_E_bous}
\end{table}
As a final check, Table \ref{tab:domain_truncation} reports the sensitivity
of $E_1^{\mathrm{Bous}}$ to domain truncation for a
rise time of 1 s. The results indicate that the
computational domain $(150,150)$ is sufficiently
large to approximate the theoretical setting
$\mathbb{R}^2$. Indeed, increasing the domain size
from 150 to 250 changes the error by only
approximately $8.90\times 10^{-4}\%$.

\begin{table}[ht]
    \centering
        \renewcommand{\arraystretch}{1.2}
    \begin{tabular}{|c|c|c|}
        \hline
        $(150,150)$ &(200,200) & $(250,250)$ \\ \hline
         $ 1.573025\times 10^{-2}$ & $ 1.573014 \times 10^{-2}$  & $1.573011\times 10^{-2}$ \\ \hline
    \end{tabular}
    \caption{{\footnotesize Sensitivity of the error $E_1^{\mathrm{Bous}}$ to domain truncation for a rise time of $1 \,\mathrm{s}$. Here, $\Delta x$ and $\Delta t$ are fixed to \eqref{Dx_Dt}, while $N_x $ is adjusted according to the domain size.
}}
    \label{tab:domain_truncation}
\end{table}

$\quad$\\
\noindent  {\bf \em Test 3: Validation of the theoretical results of Section~\ref{shalow} \vspace*{0.25cm}}
$\quad$\\
In the final test, we numerically analyze the shallow-water systems \eqref{pd1-S3} and \eqref{pd2-S3} using the same approach as the previous test, aiming to validate Theorem \ref{convergence-NSW}.

\begin{figure}[!htb]
    \centering
    \includegraphics[width=0.43\textwidth]{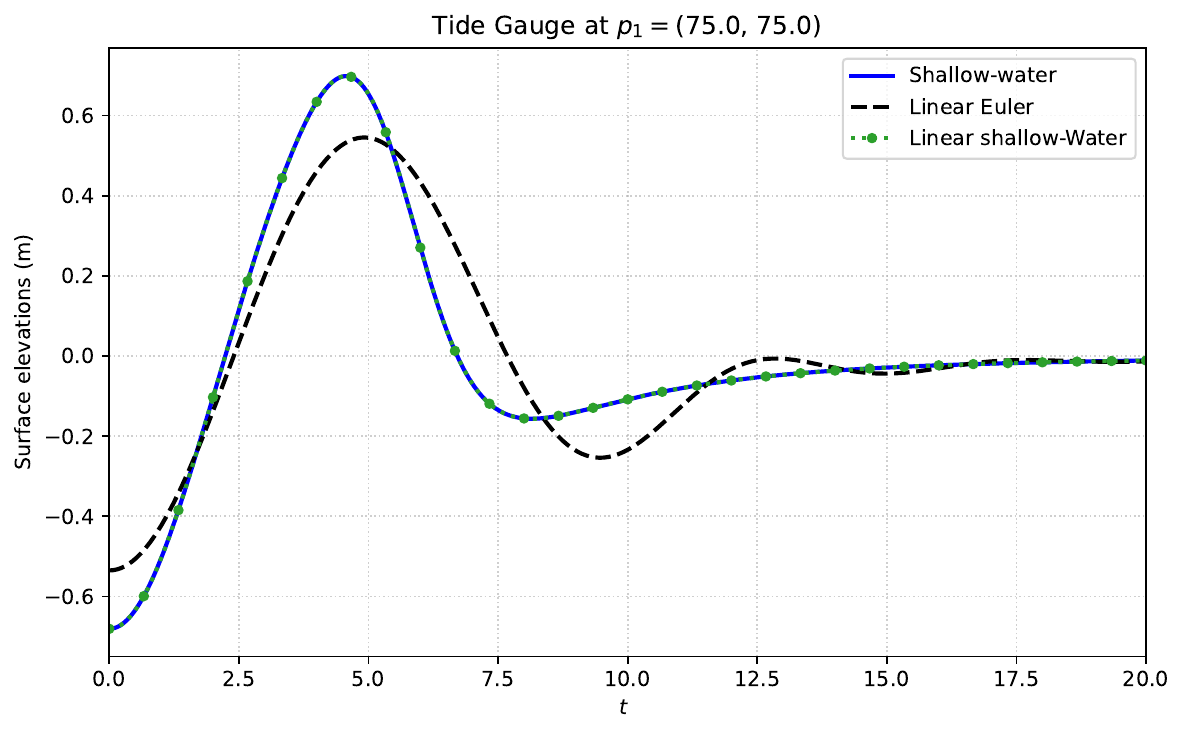}\hspace{1em}%
    \includegraphics[width=0.43\textwidth]{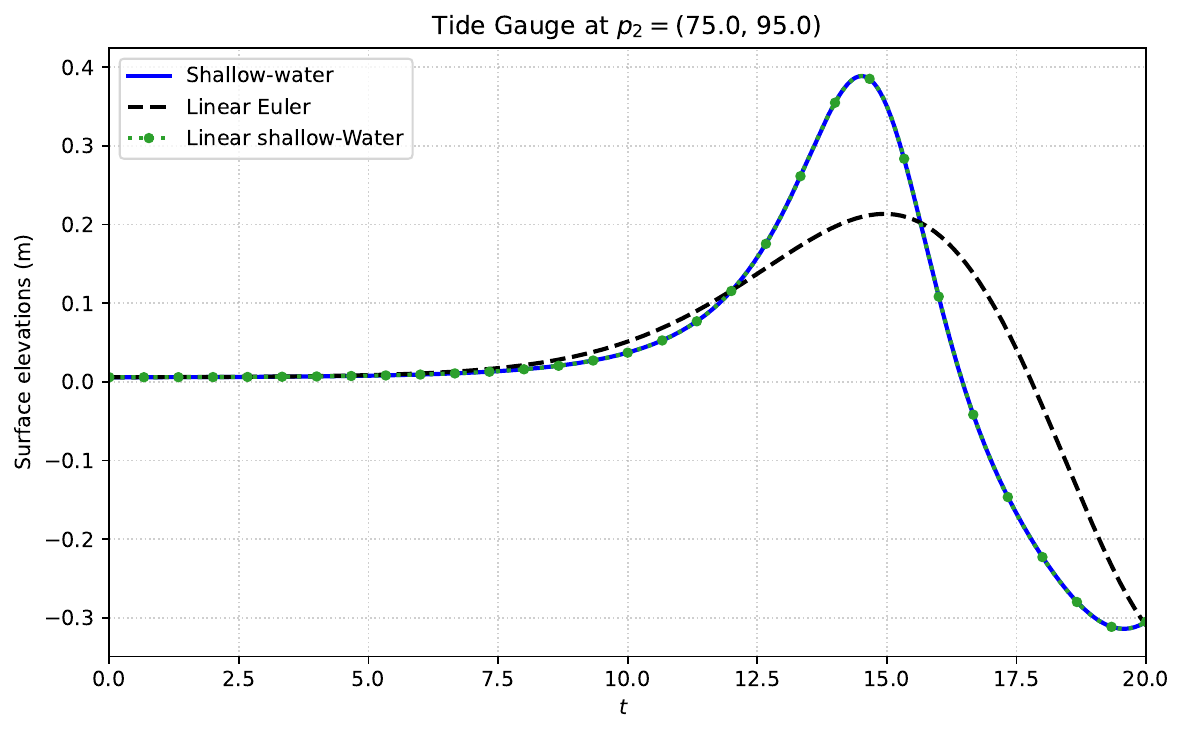}
    
    \vspace{0.5em} 
    
    \includegraphics[width=0.43\textwidth]{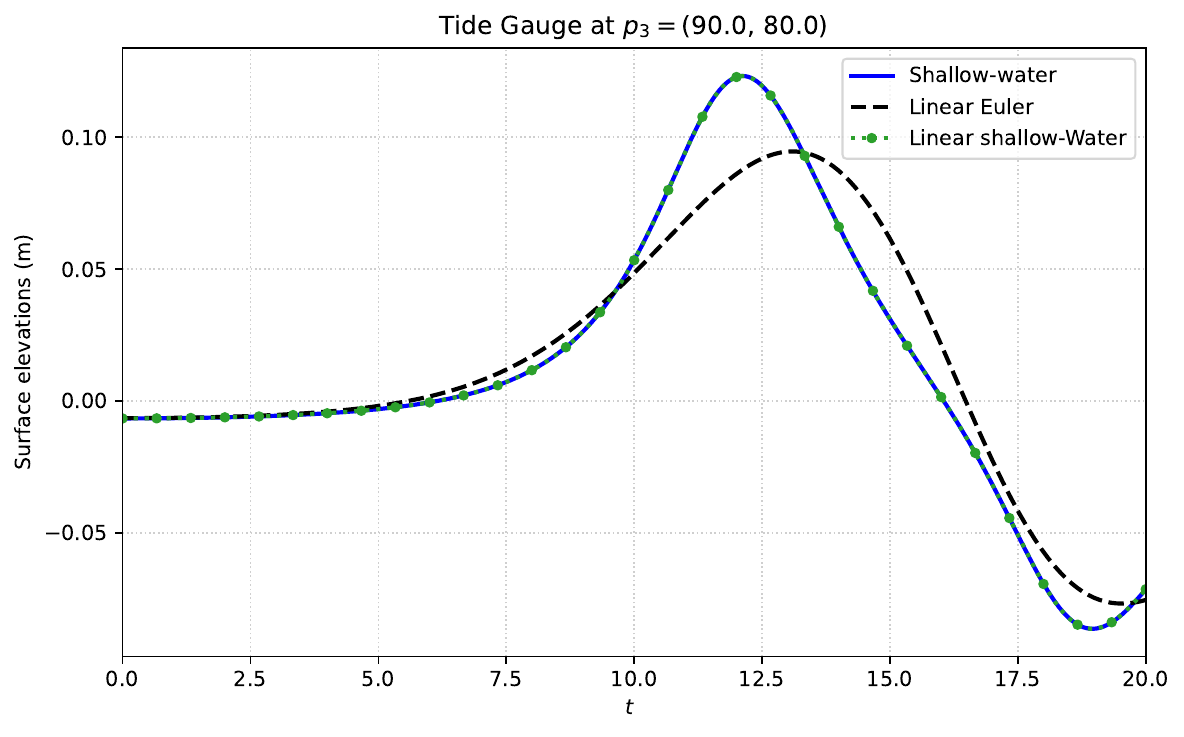}\hspace{1em}%
    \includegraphics[width=0.43\textwidth]{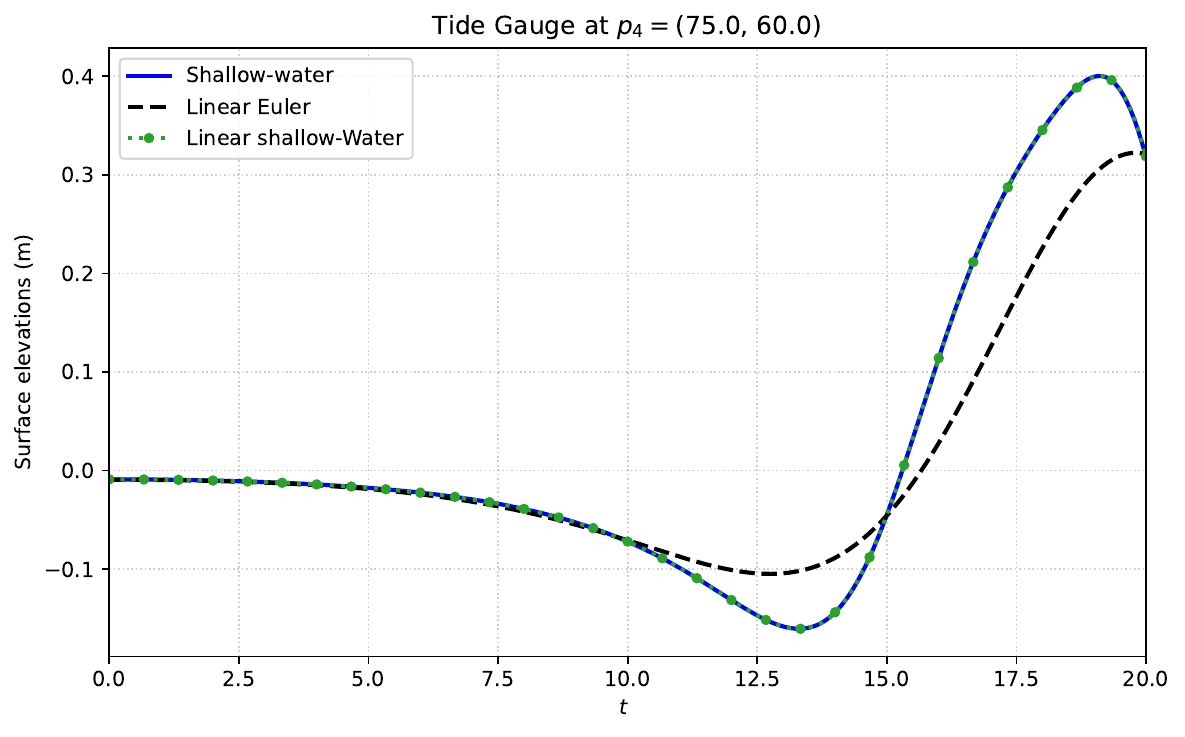}
    
    \caption{{\footnotesize Comparison of the static shallow-water,
linear Euler, and linear shallow-water solutions
at the selected tide-gauge locations.}}
    \label{fig:tide_gauges_all_SW}
\end{figure}

For comparison with linear potential theory, Figure \ref{fig:tide_gauges_all_SW}  presents the evolution of tsunami waves over time at the  locations $\{p_i\}_{1\leq i\leq 4}$, generated by the static shallow-water system and the linear Euler analytic solution \eqref{analytic_euler_solution}. In order to further validate our algorithm \eqref{General_form_discr_time}--\eqref{Elliptic_boundary}, we also compute the analytical solution of the linear static shallow-water system \eqref{pd1-S3-linear_solutions}. 

As can be clearly observed in Figure  \ref{fig:tide_gauges_all_SW}, the solution computed with our algorithm is consistent with the linear solution  \eqref{pd1-S3-linear_solutions}. However, the amplitudes of these waves are significantly greater than those predicted by both the linear Euler solution \eqref{analytic_euler_solution} and the static Boussinesq system (see Test 1),  which particularly highlights the dispersive effect discussed in Section \ref{models_dis}. We refer the reader to \cite{nhess-13-1507-2013,saito2020tsunami,zhou2012} for further details.

At the selected locations, our numerical simulations (omitted here for brevity) indicate that the kinematic shallow-water system converges to the static system similarly to the Boussinesq case in Figure~\ref{fig:convergenceP1}. We now focus on the local relative error \eqref{local_error} and the relative error \eqref{relativeerror}.

\begin{figure}[ht]
    \centering
    \includegraphics[width=0.45\textwidth]{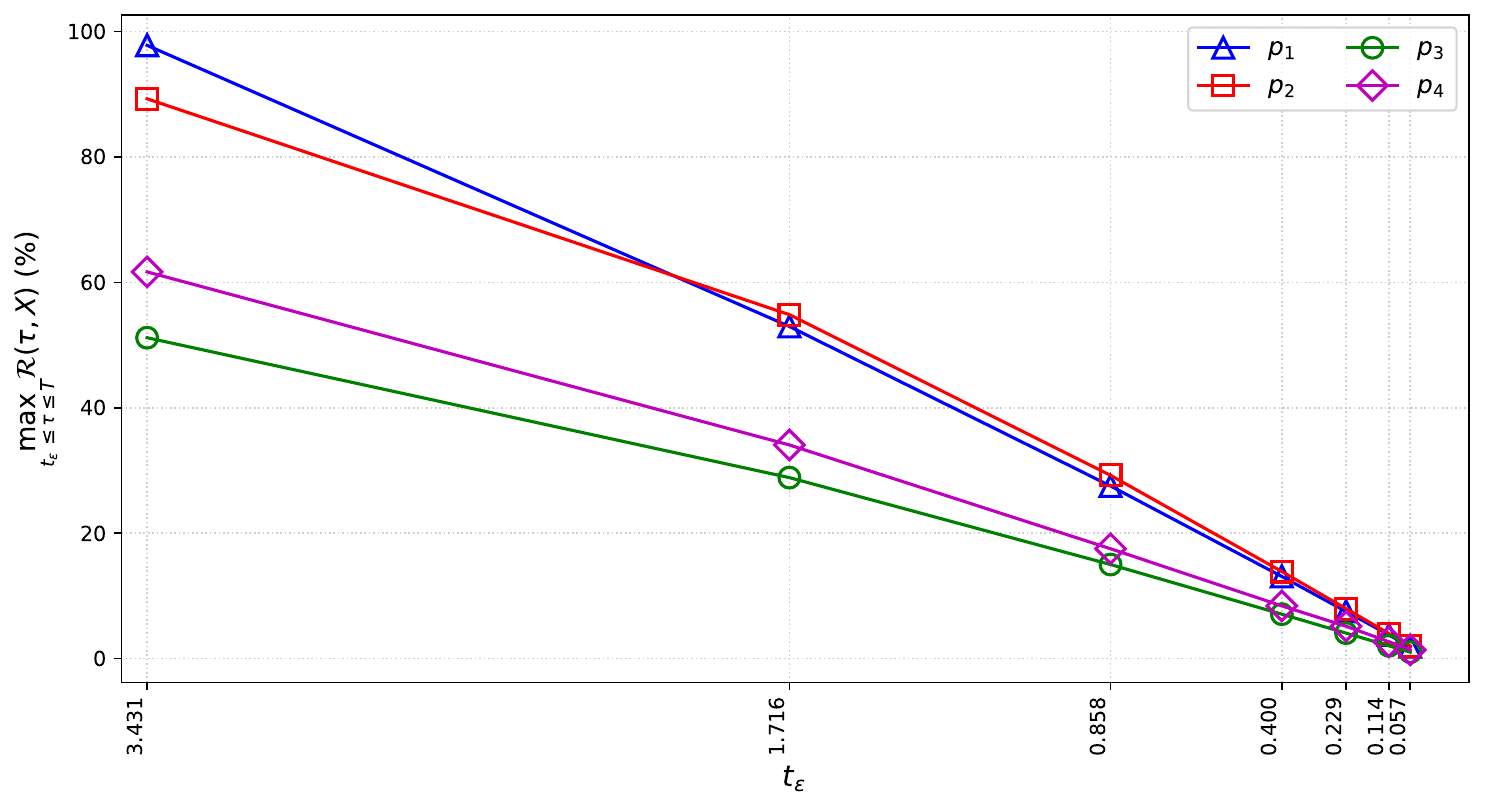}\hspace{1em}%
    \includegraphics[width=0.45\textwidth]{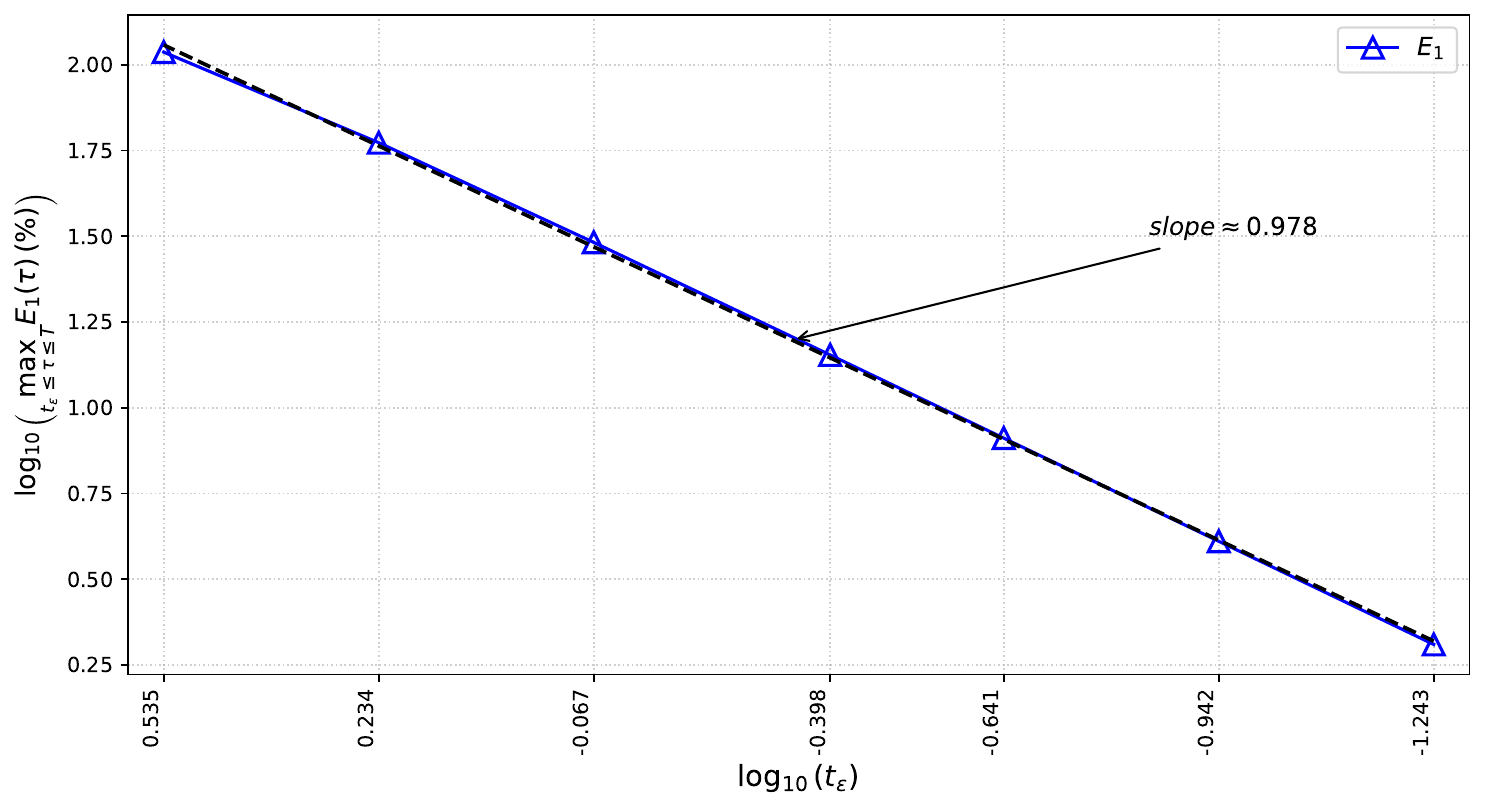}

    \caption{{\footnotesize Left: Maximum pointwise relative error $\mathcal{R}$ at selected locations $p_i$ versus the earthquake rise time $t_\epsilon$. Right: $\log_{10}$ of  the maximum  shallow-water error $E_1$ versus $\log_{10}(t_\epsilon)$. In both panels the maximum is computed over $\tau \in (t_\epsilon, 20)$.}}
    \label{fig:relativeerrorbous_NSW}
\end{figure}

In Figure \ref{fig:relativeerrorbous_NSW}, we observe similar behaviors to the Boussinesq case presented in Figure \ref{fig:relativeerrorbous}. Importantly, the error $E_1$ for $s=1$ decays almost linearly, which agrees with the theoretical estimate \eqref{after-ert-upper-bound-error-NSW}. In Table~\ref{tab:dx_dt_domain_truncation}, we report the sensitivity of the computed error $E_1$  to time and space refinement, as well as to domain truncation. As in Test~2, $E_1$ remains essentially unchanged, supporting the robustness of the 
error calculated in the right panel of
Figure~\ref{fig:relativeerrorbous_NSW}.

\begin{table}[H]
\centering
\renewcommand{\arraystretch}{1.1}
\setlength{\tabcolsep}{4pt}          

\begin{minipage}{0.50\textwidth}
    \centering
    \small                            
    \begin{tabular}{|c|c|c|c|}
        \hline
        \diagbox{$N_x$}{$N_t$} & $1000$ & $3000$ & $6000$ \\ \hline
        $200$ & $2.0481\times 10^{-2}$ & $2.0380\times 10^{-2}$ & $2.0375\times 10^{-2}$ \\ \hline
        $400$ & $2.0503\times 10^{-2}$ & $2.0402\times 10^{-2}$ & $2.0397\times 10^{-2}$ \\ \hline
        $800$ & $2.0504\times 10^{-2}$ & $2.0403\times 10^{-2}$ & $2.0398\times 10^{-2}$ \\ \hline
    \end{tabular}
\end{minipage}%
\hfill
\begin{minipage}{0.48\textwidth}
    \centering
    \small                            
    \begin{tabular}{|c|c|c|}
        \hline
        $(150,150)$ & $(200,200)$ & $(250,250)$ \\ \hline
        $ 2.040322\times 10^{-2}$ & $ 2.040307\times 10^{-2}$ & $2.040300\times 10^{-2}$ \\ \hline
    \end{tabular}
\end{minipage}

\caption{\footnotesize Sensitivity of the error $E_1$
shown in Figure~\ref{fig:relativeerrorbous_NSW} for a rise time
$t_\epsilon$ of $1\,\mathrm{s}$. Left: sensitivity to temporal
($\Delta t = 20/N_t$) and spatial refinement.
Right: sensitivity to domain truncation.}
\label{tab:dx_dt_domain_truncation}
\end{table}

\section{Conclusions, further comments, and future work}\label{sec:conclusions}
Using the shallow-water model \eqref{pd1} and a second-order dispersive 
$abcd$ model \eqref{pd2} for tsunami generation and propagation, we have derived 
upper-bound estimates for the error between the corresponding kinematic and 
static systems. These estimates depend explicitly on the earthquake rise time 
$t_\epsilon$. Consequently, in the limit $t_\epsilon \rightarrow 0$, the 
kinematic and static solutions coincide for each model. Importantly,  for the Boussinesq model, the appropriate initial free-surface condition required for this convergence is obtained by applying a Boussinesq filtering operator to the seabed \eqref{correct-intial-bousinsq}, rather than using the unfiltered bottom profile employed in the passive approach.

We validated the theoretical findings through three numerical tests using a vertical dip-slip fault as the prescribed seabed  deformation. The numerical results confirm the theoretical predictions: for  sufficiently small rise times (e.g., $1$~s), the relative errors 
remain small and the generated waves are visually indistinguishable in the plotted
comparisons. Conversely, increasing $t_\epsilon$ leads to an 
almost linear growth in the error, accompanied by a noticeable delay in tsunami 
arrival time. Furthermore, for the configurations considered, both the
static shallow-water and non-filtered static Boussinesq
systems (passive approach) overestimate the wave amplitudes
relative to the linear Euler reference solution.

We emphasize that our analysis does not provide a uniform approximation of the Euler equations as $t_\epsilon \to 0$. More precisely, the question of whether the limits $\sigma \to 0$ and $t_\epsilon \to 0$ commute is beyond the scope of the present work. We refer the reader to, e.g.,~\cite{fujiwara2015shallow,iguchi2011mathematical,lannes2013water}, for related investigations.
 
Our analysis assumes that the bottom motion is separable \eqref{bottom-type}. A natural extension is to generalize this study to non-separable bottom motions, such as those considered in \cite{sementsov2025effect}, and to multiple faults \cite{dutykh2013finite}. In addition, while the Boussinesq model considered here agrees well with 
linear potential theory when properly initialized, it would be of significant interest to derive the appropriate initial condition and an upper bound on the error for the kinematic–static comparison in the context of both the optimal dispersive 
$abcd$-Boussinesq model and the BBM-BBM model \cite{chen2003equations,mitsotakis2009boussinesq}.

\section*{Acknowledgements}
N. Lamsahel would like to express his gratitude to the Eastern Institute of Technology (EIT) for funding his postdoctoral research.

\section*{Appendix}
\appendix
\section{Proof of Theorem \ref{uniqtheorem}}\label{sec:A}
        For the sake of brevity, we will only prove the theorem for system \eqref{Bous-Pd1}; the result for system \eqref{Bous-Pd2} follows by similar arguments. For clarity, we omit the $\epsilon$-dependence in \eqref{Bous-Pd1}. Moreover, throughout the proof, we use the notation \eqref{notation}. We also set
    $$X:=L^{\infty}\left(0,T;H^{s+1} \times H^s\right),\quad Y:=C\left([0,T];H^{s}\times H^{s-1}\right), $$
    Let $s>2$ be fixed and assume that $B\in H^{s+1}$.  We introduce the  following metric space $(S_{R,T},d)$:
     \begin{equation}\label{space-for-uniqness}
         S_{R,T} :=\left\{W=(\zeta,U),\;\; W\in  X\cap Y;\quad
         \left\Vert (\zeta,U)\right\Vert_{X}\leq R \right\},
     \end{equation}
     and $$d(W_1,W_2):=\left\Vert W_1-W_2\right\Vert_{Y},$$
where $T>0$ and $R>0$ are constants to be determined presently. We can check that $(S_{R,T},d)$ is a complete metric space. Now let $\widetilde{W}\in S_{R,T}$ be given, and define the mapping $\mathcal{T}=(\mathcal{T}_1,\mathcal{T}_2)$ by 
\begin{equation}\label{mapping}
	\left\{
	\begin{array}{ll}
		\zeta(X,t)=\mathcal{T}_1(\widetilde{W})=-\displaystyle\int_0^t\left[\left(I-\frac{1}{3}\Delta\right)^{-1}\,\nabla\cdot\left(\widetilde{H}\,\widetilde{U}\right)\right](X,\tau)\,\dd \tau\;\;+ \displaystyle\int_0^t\left[\left(I-\frac{1}{3}\Delta\right)^{-1}(I+\frac{1}{6}\Delta )b_t\right](X,\tau)\,\dd \tau,\vspace{0.05cm}\\
        U(X,t)=\mathcal{T}_2(\widetilde{W}),
	\end{array}
	\right.
\end{equation}
where  $U$ is defined as the solution of the following linear transport problem:
\begin{equation}\label{mapping2}
	\left\{
	\begin{array}{ll}
		U_t+ (\widetilde{U}.\nabla)U=-\nabla\widetilde{\zeta},\vspace{0.05cm}\\
        U(X,0)=\mathbf{0}.
	\end{array}
	\right.
\end{equation}
The above mapping is well defined. Indeed, we infer from \cite[Theorem~3.19]{bahouri2011fourier} that system \eqref{mapping2} admits a unique solution $U$ in $C([0,T],H^s).$ In the next steps of the proof, we will show that $T$ and $R$ can be chosen such  that the mapping $\mathcal{T}$ is a contraction on $S_{R,T}$.\\
Using the algebra property of \(H^r\), \(r>1\), together with the boundedness of the operator
\begin{equation}\label{oper-A}
A=\left(I-\frac{1}{3}\Delta\right)^{-1}\nabla\cdot
\colon H^s\longrightarrow H^{s+1},    
\end{equation}
whose operator norm is bounded by $3$, we obtain
\begin{equation*}
	\begin{split}
		\left\Vert \mathcal{T}_1(\widetilde{W})\right\Vert_{H^{s+1}}&\leq f_\epsilon(t)\left\Vert\left(I-\frac{1}{3}\Delta\right)^{-1} (I+\frac{1}{6}\Delta ) B \right\Vert_{H^{s+1}}+ \displaystyle\int_0^t\left\Vert A\left(\widetilde{H}\,\widetilde{U}\right) (\tau)\right\Vert_{H^{s+1}}\,\dd \tau  \\
        &\leq\left\Vert B\right\Vert_{H^{s+1}}+3\displaystyle\int_0^t\left\Vert\widetilde{H}\,\widetilde{U}\right\Vert_{H^{s}}(\tau)\,\dd \tau \\
        &\leq  \left\Vert B\right\Vert_{H^{s+1}}+C_s\displaystyle\int_0^t \left(1+\left\Vert \widetilde{\zeta}\right\Vert_{H^s} +\left\Vert b\right\Vert_{H^s}   \right)\left\Vert \widetilde{U}\right\Vert_{H^s}(\tau)\,\dd \tau\\
        &\leq  \left\Vert B\right\Vert_{H^{s+1}}+C_s T\left( 1+\left\Vert \widetilde{\zeta} \right\Vert_{L^{\infty}(0,T;H^{s+1})}+\left\Vert B\right\Vert_{H^{s}}\right)\left\Vert \widetilde{U} \right\Vert_{L^{\infty}(0,T;H^{s})}\\
        &\leq  \left\Vert B\right\Vert_{H^{s+1}}+C_s T(1+R+\left\Vert B\right\Vert_{H^{s+1}})R.
	\end{split}
\end{equation*}
On the other hand, we infer from \cite[Theorem~3.14]{bahouri2011fourier} that there exists a constant $C_1(s)>0$, depending only on $s$, such that, for a.e. $ t\in(0,T)$, we have 
\begin{equation*}
	\left\Vert U(t)\right\Vert_{H^s}\leq  \,e^{C_1(s)\displaystyle\int_0^t\left\Vert\widetilde{U}(\tau )\right\Vert_{H^s}\,\dd \tau } \displaystyle\int_0^t \left\Vert\nabla \widetilde{\zeta}\right\Vert_{H^s}(\tau)\,\dd \tau.
\end{equation*}
Hence,
\begin{equation*}
	\begin{split}
		\left\Vert \mathcal{T}_2(\widetilde{W})\right\Vert_{H^s}\leq  \,T R e^{C_1(s) T R}.
	\end{split}
\end{equation*}
By choosing $R= 4\left\Vert B\right\Vert_{H^{s+1}}$ and  
 $$T\leq\sup \left\{t\in \left(0, \frac{1}{ C_s(4+5R)}\right);\quad t \,e^{C_1(s)R\,t} -\frac{1}{2}\leq 0\right\},$$
we get $\mathcal{T}( S_{R,T})\subseteq S_{R,T} $. It remains to prove that  $\mathcal{T}$ is a contraction.  Let us fix $\widetilde{W_1}, \widetilde{W_2}$ in $S_{R,T}$. We have
\begin{equation}\label{contarction1}
	\begin{split}
		\left\Vert\mathcal{T}_1(\widetilde{W_1})-\mathcal{T}_1(\widetilde{W_2})\right\Vert_{H^s}&\leq \displaystyle\int_0^t \left \Vert A \left((\widetilde{U_1}-\widetilde{U_2})-b(\widetilde{U_1}-\widetilde{U_2})+\widetilde{\zeta_1}\widetilde{U_1}-\widetilde{\zeta_2}\widetilde{U_2}\right)\right\Vert_{H^s}(\tau)\,\dd \tau\\
        &\leq C_s\displaystyle\int_0^t \Bigl[\left\Vert\widetilde{U_1}-\widetilde{U_2}\right\Vert_{H^{s-1}}+\left\Vert b\right\Vert_{H^{s-1}}\left\Vert\widetilde{U_1}-\widetilde{U_2}\right\Vert_{H^{s-1}}+\left\Vert\widetilde{\zeta_1}\right\Vert_{H^{s-1}}\left\Vert\widetilde{U_1}-\widetilde{U_2}\right\Vert_{H^{s-1}}\\&\quad\quad\quad\quad\quad+\left\Vert\widetilde{U_2}\right\Vert_{H^{s-1}}\left\Vert\widetilde{\zeta_1}-\widetilde{\zeta_2}\right\Vert_{H^{s-1}}\Bigr](\tau)\;\dd \tau\\
        &\leq T C_s\frac{4+9R}{4} \left\Vert\widetilde{W_1}-\widetilde{W_2}\right\Vert_{Y}.
	\end{split} 
\end{equation}
On the other hand, the difference $\mathcal{T}_2(\widetilde{W_1})-\mathcal{T}_2(\widetilde{W_2}):=U_1-U_2$ is the unique solution of the following transport system:
\begin{equation}\label{U1U2-mapping}
	\left\{
	\begin{array}{ll}
		(U_1-U_2)_t+ (\widetilde{U}_2.\nabla)(U_1-U_2)=-\,\nabla\left(\widetilde{\zeta_1}-\widetilde{\zeta_2}\right)-\left( \left( \widetilde{U}_1- \widetilde{U}_2\right)\cdot\nabla \right)U_1,\vspace{0.05cm}\\
        (U_1-U_2)(X,0)=\mathbf{0}.
	\end{array}
	\right.
\end{equation}
Applying again \cite[Theorem~3.14]{bahouri2011fourier}, we obtain

\begin{equation}\label{contarction2}
\begin{split}
	\left\Vert (U_1-U_2)(t)\right\Vert_{H^{s-1}}&\leq  \,e^{C_1(s)\displaystyle\int_0^t\left\Vert\widetilde{U_2}(\tau )\right\Vert_{H^{s}}\,\dd \tau } \displaystyle\int_0^t \left[\left\Vert\widetilde{\zeta}_1-\widetilde{\zeta}_2\right\Vert_{H^{s}} + C_s\left\Vert\widetilde{U}_1-\widetilde{U}_2\right\Vert_{H^{s-1}}\left\Vert U_1\right\Vert_{H^{s}}\right](\tau)\,\dd \tau\\
    &\leq \, e^{C_1(s)TR} \,T\left(1+C_sR\right)\left\Vert\widetilde{W_1}-\widetilde{W_2}\right\Vert_{Y}.
    \end{split}
\end{equation}
Adding \eqref{contarction1} and \eqref{contarction2} term by term, we obtain
$$d\left(\mathcal{T}\left(\widetilde{W_1}\right),\mathcal{T}\left(\widetilde{W_2}\right)\right)\leq L_{\mathcal{T}}\;d\left(\widetilde{W_1},\widetilde{W_2}\right),$$
where $$L_{\mathcal{T}} =2\max\left\{C_s T \frac{4+9R}{4},  \,e^{C_1(s)TR} \,T\left(1+C_sR\right)\right\}.$$
Therefore, to ensure that the mapping is a contraction ($L_{\mathcal{T}}\leq 1/2$) and maps $S_{R,T}$ into itself, we choose $T$ as follows:
$$T=T^{\star}:=\sup\left\{t\in \left(0, \frac{1}{ C_s(4+9R)}\right);\quad t \,e^{C_1(s)R\,t} -\frac{1}{4(1+C_s R)}\leq 0\right\}.$$
By the Banach fixed-point theorem, the mapping $\mathcal{T}$ admits a unique fixed point $(\underline{\zeta},\underline{U})\in S_{R,T}$. Therefore, the fixed point $(\underline{\zeta},\underline{U})$ is the unique solution of \eqref{Bous-Pd1}. This completes the proof.



\begin{thebibliography}{10}










\bibitem{bahouri2011fourier}
H.~Bahouri, J.-Y.~Chemin, R.~Danchin.
\newblock {\em Fourier analysis and nonlinear partial differential equations}, volume~343.
\newblock Springer, Heidelberg, 2011.

\bibitem{bona1998boussinesq}
J.~L.~Bona, M.~Chen.
\newblock {\em A Boussinesq system for two-way propagation of nonlinear dispersive waves}.
\newblock Physica D: Nonlinear Phenomena, 116(1):191--224, 1998.

\bibitem{bona2002boussinesq}
J.~L.~Bona, M.~Chen, J.-C.~Saut.
\newblock {\em Boussinesq equations and other systems for small-amplitude long waves in nonlinear dispersive media. I: Derivation and linear theory}.
\newblock Journal of Nonlinear Science, 12:283--318, 2002.


\bibitem{bona2004}
J.~L.~Bona, M.~Chen,  J.-C.~Saut.
\newblock {\em Boussinesq equations and other systems for small-amplitude long waves in nonlinear dispersive media: II. The nonlinear theory}.
\newblock Nonlinearity, 17(3):925--952, 2004.







\bibitem{chen2003equations}
M.~Chen.
\newblock {\em Equations for bi-directional waves over an uneven bottom}.
\newblock Mathematics and Computers in Simulation, 62(1):3--9, 2003.

\bibitem{chen2009numerical}
M.~Chen.
\newblock {\em Numerical investigation of a two-dimensional Boussinesq system}.
\newblock Discrete and Continuous Dynamical Systems, 23(4):1169--1190, 2009.

\bibitem{chen2020mathematical}
M.~Chen, O.~Goubet, S.~Li.
\newblock {\em Mathematical analysis of bump to bucket problem}.
\newblock Communications on Pure and Applied Analysis, 19(12):5567--5580, 2020.



\bibitem{coulaud2025comparison}
G.~Coulaud, M.~Teles,   M.~Benoit.
\newblock {\em A comparison of eight weakly dispersive Boussinesq-type models for non-breaking long-wave propagation in variable water depth}.
\newblock Coastal Engineering, 195:104645, 2025.








\bibitem{stefano2000}
F.~Di Benedetto, S.~Serra-Capizzano.
\newblock {\em Optimal multilevel matrix algebra operators}.
\newblock Linear and Multilinear Algebra, 48:35–66, 2000.



\bibitem{dutykh2006linear}
D.~Dutykh, F.~Dias, Y.~Kervella.
\newblock {\em Linear theory of wave generation by a moving bottom}.
\newblock Comptes Rendus Mathematique, 343(7):499--504, 2006.

\bibitem{dutykh2007water}
D.~Dutykh, F.~Dias.
\newblock {\em Water waves generated by a moving bottom}.
\newblock In A.~Kundu (ed.), {\em Tsunami and Nonlinear Waves}, pages~65--95, Springer, Berlin, Heidelberg, 2007.


\bibitem{dutykh2009tsunami}
D.~Dutykh, F.~Dias.
\newblock {\em Tsunami generation by dynamic displacement of sea bed due to dip-slip faulting}.
\newblock Mathematics and Computers in Simulation, 80(4):837--848, 2009.


\bibitem{dutykh2011}
D.~Dutykh, R.~Poncet, F.~Dias.
\newblock {\em The VOLNA code for the numerical modeling of tsunami waves: Generation, propagation and inundation}.
\newblock European Journal of Mechanics-B/Fluids, 30(6):598--615, 2011.


\bibitem{dutykh2013finite}
D.~Dutykh, D.~Mitsotakis, X.~Gardeil, F.~Dias.
\newblock {\em On the use of the finite fault solution for tsunami generation problems}.
\newblock Theoretical and Computational Fluid Dynamics, 27:177--199, 2013.




\bibitem{fontelos2017bottom}
M.~A.~Fontelos, R.~Lecaros, J.~C.~L\'opez-R\'{\i}os, J.~H.~Ortega.
\newblock {\em Bottom detection through surface measurements on water waves}.
\newblock SIAM Journal on Control and Optimization, 55(6):3890--3907, 2017.

\bibitem{fujiwara2015shallow}
H.~Fujiwara, T.~Iguchi.
\newblock {\em A shallow-water approximation for water waves over a moving bottom}.
\newblock Nonlinear Dynamics in Partial Differential Equations, 64:77--88, 2015.







\bibitem{nhess-13-1507-2013}
S.~Glimsdal, G.~K.~Pedersen, C.~B.~Harbitz, F.~L{\o}vholt.
\newblock {\em Dispersion of tsunamis: does it really matter?}.
\newblock Natural Hazards and Earth System Sciences, 13(6):1507--1526, 2013.


\bibitem{grilli1994shoaling}
S. T.~Grilli, R.~Subramanya, I. A.~Svendsen, J.~Veeramony.
\newblock {\em Shoaling of solitary waves on plane beaches.}
\newblock  Journal of Waterway, Port, Coastal, and Ocean Engineering, 120:609--628, 1994.

\bibitem{grilli2007source}
S.~T.~Grilli, M.~Ioualalen, J.~Asavanant, F.~Shi, J.~T.~Kirby, P.~Watts.
\newblock {\em Source constraints and model simulation of the December 26, 2004, Indian Ocean tsunami}.
\newblock Journal of Waterway, Port, Coastal, and Ocean Engineering, 
133:414--428, 2007.


\bibitem{hammack1973}
J.~L.~Hammack.
\newblock {\em A note on tsunamis: their generation and propagation in an ocean of uniform depth}.
\newblock Journal of Fluid Mechanics, 60(4):769--799, 1973.




\bibitem{iguchi2011mathematical}
T.~Iguchi.
\newblock {\em A mathematical analysis of tsunami generation in shallow-water due to seabed deformation}.
\newblock Proceedings of the Royal Society of Edinburgh: Section A Mathematics, 141(3):551--608, 2011.

\bibitem{imamura2019}
F.~Imamura, S.~P.~Boret, A.~Suppasri, A.~Muhari.
\newblock {\em Recent occurrences of serious tsunami damage and the future challenges of tsunami disaster risk reduction}.
\newblock Progress in Disaster Science, 1:100009, 2019.



\bibitem{kajiura1963leading}
K.~Kajiura.
\newblock {\em The leading wave of a tsunami}.
\newblock Bulletin of the Earthquake Research Institute, 41:535--571, 1963.


\bibitem{kervella2007}
Y.~Kervella, D.~Dutykh, F.~Dias.
\newblock {\em Comparison between three-dimensional linear and nonlinear tsunami generation models}.
\newblock Theoretical and Computational Fluid Dynamics, 21:245--269, 2007.

\bibitem{kubota2026}
T.~Kubota, T.~Saito.
\newblock {\em Does three‐dimensional structure affect tsunami excitation and stress drop estimates?—The case of the 2011 Tohoku earthquake}.
\newblock Geophysical Research Letters, 53:e2026GL122518, 2026.


\bibitem{lamsahel2025outlier}
N.~Lamsahel, C.~Manni, A.~Ratnani, S.~Serra-Capizzano, H.~Speleers.
\newblock {\em Outlier-free isogeometric discretizations for Laplace eigenvalue problems: closed-form eigenvalue and eigenvector expressions}.
\newblock Numerische Mathematik, 157(4):1397--1448, 2025.


\bibitem{lamsahel2026uniqueness}
N.~Lamsahel, L.~Rosier.
\newblock {\em Uniqueness and stability in bottom detection through surface measurements of water waves}.
\newblock  arXiv:2601.17639, 2026.

\bibitem{lannes2013water}
D.~Lannes.
\newblock {\em The water-waves problem: mathematical analysis and asymptotics}, volume~188.
\newblock American Mathematical Society, Providence RI, 2013.


\bibitem{lecaros2024optimal}
R.~Lecaros, J.~L{\'o}pez-R{\'\i}os, G.~I.~Montecinos, E.~Zuazua.
\newblock {\em Optimal control approach for moving bottom detection in one-dimensional shallow-waters by surface measurements}.
\newblock Mathematical Methods in the Applied Sciences, 47:13973--14004, 2024.

\bibitem{levin2009}
B.~Levin, M.~Nosov.
\newblock {\em Physics of tsunamis}.
\newblock Springer Netherlands, Dordrecht, 2009.



\bibitem{lotto2017}
G.~C.~Lotto, G.~Nava, E.~M.~Dunham.
\newblock {\em Should tsunami simulations include a nonzero initial horizontal velocity?}.
\newblock Earth, Planets and Space, 69:117, 2017.

\bibitem{lovholt2012}
F.~Løvholt, G.~Pedersen, S.~Bazin,  D. Kühn, R.~E.~Bredesen, C.~Harbitz.
\newblock {\em Stochastic analysis of tsunami run-up due to heterogeneous co-seismic slip and dispersion}.
\newblock Journal of Geophysical Research, 117:C03047, 2012.


\bibitem{lucquin2004}
B.~Lucquin.
\newblock {\em Équations aux dérivées partielles et leurs approximations}.
\newblock Ellipses, Paris, 2004.








\bibitem{MELKIOR2026114990}
T.~Melkior, H.~S.~Bhat, F.~Amlani.
\newblock {\em Tsunami modeling with dynamic seafloors: A high-order solver validated with shallow-water benchmarks}.
\newblock Journal of Computational Physics, 562:114990, 2026.


\bibitem{mitsotakis2009boussinesq}
D.~E.~Mitsotakis.
\newblock {\em Boussinesq systems in two space dimensions over a variable bottom for the generation and propagation of tsunami waves}.
\newblock Mathematics and Computers in Simulation, 80(4):860--873, 2009.





\bibitem{nosov1996}
M.~A.~Nosov.
\newblock {\em A comparative study of tsunami excited by piston-type and traveling-wave bottom motion}.
\newblock Volcanology and Seismology, 6:693--698, 1996.




\bibitem{ohmachi2001}
T.~Ohmachi, H.~Tsukiyama, H.~Matsumoto.
\newblock {\em Simulation of tsunami induced by dynamic displacement of seabed due to seismic faulting}.
\newblock Bulletin of the Seismological Society of America, 91(6):1898--1909, 2001.

\bibitem{okada1985surface}
Y.~Okada.
\newblock {\em Surface deformation due to shear and tensile faults in a half-space}.
\newblock Bulletin of the Seismological Society of America, 75(4):1135--1154, 1985.

\bibitem{okada1992internal}
Y.~Okada.
\newblock {\em Internal deformation due to shear and tensile faults in a half-space}.
\newblock Bulletin of the Seismological Society of America, 82(2):1018--1040, 1992.

\bibitem{outiskt2026}
M.~Outiskt, S.~Tichli, M.~A.~Baptista, S.~El Moussaoui, K.~Aboumaria.
\newblock {\em Predictive study of tsunami impact on the southwestern Atlantic coast of Morocco}.
\newblock Journal of African Earth Sciences, 236:106025, 2026.





\bibitem{peregrine1967long}
D. H.~Peregrine.
\newblock {\em Long waves on a beach.}
\newblock Journal of Fluid Mechanics, 27(4):815--827, 1967.

\bibitem{poletto2025full}
J.~V.~P.~Poletto, D.~Andrade, M.~V.~Flamarion, R.~Ribeiro-Jr.
\newblock {\em Full Euler equations for waves generated by vertical seabed displacements}.
\newblock SIAM Journal on Applied Mathematics, 85(4):1344--1360, 2025.




\bibitem{saillour2022modelling}
T.~Saillour, G.~Cozzuto, G.~Lupoi, G.~Cuomo, S.~E.~Bourban.
\newblock {\em Modelling historical global tsunamis with TELEMAC}.
\newblock 28th TELEMAC User Conference, 2022.


\bibitem{saito2010}
T.~Saito, K.~Satake, T.~Furumura.
\newblock {\em Tsunami waveform inversion including dispersive waves: the 2004 earthquake off Kii Peninsula, Japan}.
\newblock Journal of Geophysical Research, 115:B06303, 2010.

\bibitem{saito2013}
T.~Saito.
\newblock {\em Dynamic tsunami generation due to sea-bottom deformation: Analytical representation based on linear potential theory}.
\newblock Earth Planets and Space, 65:1411--1423, 2013.

\bibitem{saito2019tsunami}
T.~Saito.
\newblock {\em Tsunami Generation and Propagation}.
\newblock Springer, Tokyo, 2019.

\bibitem{saito2020tsunami}
T.~Saito, T.~Kubota.
\newblock {\em Tsunami modeling for the deep sea and inside focal areas}.
\newblock Annual Review of Earth and Planetary Sciences, 48:121--145, 2020.



\bibitem{sementsov2025effect}
K.~A.~Sementsov, T.~Baba, S.~V.~Kolesov, Y.~Tanioka, M.~A.~Nosov.
\newblock {\em The effect of earthquake fault rupture kinematics on tsunami generation: a numerical study of real events}.
\newblock Geophysical Journal International, 240:920--941, 2025.


\bibitem{shen1994efficient}
J.~Shen.
\newblock {\em Efficient spectral-Galerkin method I. direct solvers for second- and fourth-order equations by using Legendre polynomials}.
\newblock SIAM Journal on Scientific Computing, 15(6):1489--1505, 1994.


\bibitem{synolakis1991green}
C.~E.~Synolakis.
\newblock {\em Green's law and the evolution of solitary waves}.
\newblock Physics of Fluids A: Fluid Dynamics, 3(3):490, 1991.

\bibitem{synolakis2006}
C.~E.~Synolakis, E.~N.~Bernard.
\newblock {\em Tsunami Science before and beyond Boxing Day 2004}.
\newblock Philosophical Transactions: Mathematical, Physical and Engineering Sciences, 364(1845):2231--2265, 2006.




\bibitem{taylor1997nonlinear}
M.~E.~Taylor.
\newblock {\em Partial differential equations. III. Nonlinear Equations}, volume~117.
\newblock Springer-Verlag, New York, 1997.

\bibitem{triantafyllou2024strong}
I.~Triantafyllou, A.~Agalos, A.~G.~Samaras, T.~V.~Karambas, G.~A.~Papadopoulos.
\newblock {\em Strong earthquakes and tsunami potential in the Hellenic Subduction Zone}.
\newblock Journal of Geodynamics, 159:102021, 2024.


\bibitem{ursell1953}
F.~Ursell.
\newblock {\em The long-wave paradox in the theory of gravity waves}.
\newblock Mathematical Proceedings of the Cambridge Philosophical Society, 49(4):685--694, 1953.



\bibitem{wang2021}
Y.~Wang, H. Tsushima, K. Satake, P. Navarrete.
\newblock {\em Review on recent progress in near-field tsunami forecasting using offshore tsunami measurements: Source inversion and data assimilation}.
\newblock Pure and Applied Geophysics, 178:5109--5128, 2021.


\bibitem{zhao2024}
G.~Zhao, X.~Niu.
\newblock {\em Possible maximum earthquake tsunami along the south coast of China inferred from GPS-derived surface velocities}.
\newblock Journal of Geophysical Research: Oceans, 129:e2023JC020835, 2024.

\bibitem{zhou2012}
H.~Zhou, Y.~Wei, V.~V.~Titov.
\newblock {\em Dispersive modeling of the 2009 Samoa tsunami}.
\newblock Geophysical Research Letters, 39:L16303, 2012.

\end{thebibliography}
\end{document}